# Energy Efficiency in MIMO Underlay and Overlay Device-to-Device Communications and Cognitive Radio Systems

Alessio Zappone, *Senior Member, IEEE*, Bho Matthiesen, *Student Member, IEEE*, and Eduard A. Jorswieck, *Senior Member, IEEE*

*Abstract*—This paper addresses the problem of resource allocation for systems in which a primary and a secondary link share the available spectrum by an underlay or overlay approach. After observing that such a scenario models both cognitive radio and D2D communications, we formulate the problem as the maximization of the secondary energy efficiency subject to a minimum rate requirement for the primary user. This leads to challenging non-convex, fractional problems. In the underlay scenario, we obtain the global solution by means of a suitable reformulation. In the overlay scenario, two algorithms are proposed. The first one yields a resource allocation fulfilling the first-order optimality conditions of the resource allocation problem, by solving a sequence of easier fractional problems. The second one enjoys a weaker optimality claim, but an even lower computational complexity. Numerical results demonstrate the merits of the proposed algorithms both in terms of energy-efficient performance and complexity, also showing that the two proposed algorithms for the overlay scenario perform very similarly, despite the different complexity.

## I. Introduction

Two precious resources which are starting to become scarce in wireless communication systems are energy and spectrum.

Credited sources foresee that by 2020 the number of devices connected to the Internet will be more than 50 Billion, with more than six devices per person [1]. In order to sustain networks with such a large number of devices to serve, 5G networks will be required to achieve a 1000x increase of the data rate, compared to present wireless networks. However, it is clear that this rate growth can not be achieved by simply scaling up the transmit powers, due to both sustainable growth and environmental concerns. Already nowadays information communication technologies (ICT) infrastructures are responsible for more than 3% of the world-wide energy consumption, which causes more than 5% of the world-wide $CO_2$ emissions [2], and the situation is going to escalate given the foreseen exponential increase of connected devices. In order to stop this escalation, the 1000x data-rate increase should be obtained with an energy consumption comparable to that of present networks. This calls for a paradigm shift in the way cellular networks are designed. The main performance indicator should not only be the amount of transferred information, but rather the amount of information which can be transferred per Joule of available energy, i.e. the network bit/Joule energy efficiency. Among the different performance measures which have been proposed to capture the fundamental trade-off between transmitting at high rates and limiting energy consumptions, the most well-established one is the network global energy efficiency (GEE), defined as the ratio between the system sum-rate and consumed power [3], [4].

Besides energy, another resource which is running short in wireless communications is spectrum. It is known that the current fixed spectrum allocation policy results in underutilization of precious spectrum resources, and the spectrum demand is increasing with rapid growth of wireless applications and devices. To tackle the spectrum scarcity problem, one well-established approach is to allow an unlicensed network, termed secondary system, to reuse the spectrum owned by a legitimate, primary, system, provided this does not degrade the quality of service (QoS) of the licensed system below a predefined threshold [5], [6]. One well-established example of this idea is the cognitive radio technique [7], but more recently this approach has also been proposed for device-to-device (D2D) communications in cellular networks, which is emerging as one candidate technology for 5G networks, due to both the proximity gain that it grants, and to the possibility of performing efficient base station off-loading [8]. In both scenarios, spectrum sharing can be performed in an underlay fashion [9], in which the primary system is oblivious to the presence of the secondary system, or in an overlay fashion [10], in which primary and secondary users cooperate to achieve mutual benefit.

### A. Related literature

Most previous works in the field of cooperative resource allocation in underlay and overlay communications consider

Parts of this paper were presented at the 15th IEEE International Workshop on Signal Processing Advances in Wireless Communications, 2014, and at the 2016 IEEE Wireless Communications and Networking Conference.

A. Zappone is with the Department of Electrical and Information Engineering of the University of Cassino and Southern Lazio, Cassino, Italy, alessio.zappone@unicas.it. B. Matthiesen and E. A. Jorswieck are with the Chair for Communications Theory, Communications Laboratory, Technische Universität Dresden, Dresden, Germany (e-mail: {bho.matthiesen, eduard.jorswieck}@tu-dresden.de).

The work of A. Zappone was performed as he was with the Communications Laboratory of the Technische Universität Dresden, and was supported by the German Research Foundation (DFG), under grant ZA 747/1-3.

The work of E. Jorswieck and B. Matthiesen was supported in part by the German Research Foundation, Deutsche Forschungsgemeinschaft (DFG) in the Collaborative Research Center 912 Highly Adaptive Energy-Efficient Computing.

single-antenna systems (see [10]–[13] and references therein). However, the use of multiple antennas has been shown to allow a more effective interference management, due to the extra available spatial dimensions [14]. A non-cooperative game-theoretic approach for resource allocation in multiple-input multiple-output (MIMO) cognitive radio systems is employed in [15], where conditions for the existence and uniqueness of Nash equilibria are derived. A similar approach is used in [16], where imperfect channel state information (CSI) is assumed and a robust optimization approach is taken. In [17] a MIMO cognitive radio system with multiple secondary data streams is considered, and beamforming schemes for transmit power minimization subject to individual signal-to-noise ratio (SNR) requirements for each data stream are proposed. Further contributions in MIMO underlay cognitive radio systems are provided in [18]–[20], where different resource allocation algorithms for rate optimization are proposed. In [21] a MIMO device-to-device (D2D) underlay network is analyzed, and resource allocation algorithms for sum rate maximization are developed, using a pricing-based approach. In [22] a two-way, D2D relay-assisted MIMO network is addressed, and linear transceiver design is carried out for minimum mean square error minimization. The coexistence of MIMO and D2D underlay is addressed in [23], which however is more focused on modeling aspects rather than on radio resource allocation algorithms.

All above works consider resource allocation problems aimed at optimizing traditional performance measures like the sum rate or the minimum mean square error, which are not energy-efficient. Resource allocation for energy efficiency has been investigated in fewer contributions, mainly from a game-theoretic perspective. In [24] coalitional game theory is used to develop distributed algorithms to optimize the energy efficiency of the mobiles, while non-cooperative game theory is used in [25] to perform competitive and energy-efficient resource allocation. The use of Stackelberg games is instead proposed in [26], for a cognitive radio system in which the primary system loans spectrum to the secondary system. A relay-assisted cognitive radio system is considered in [27], studying the problem of joint relay selection and power control. A centralized approach is taken in [28] for GEE maximization in an orthogonal frequency division multiple access (OFDMA) cognitive system. Fractional programming is employed assuming no multi-user interference is present. The trade-off between energy efficiency and spectral efficiency in D2D communications has been analyzed in [29].

However, all of the mentioned works on energy efficiency focus on single-antenna systems. A first work to discuss the problem of green resource allocation in MIMO networks is [30], which, however, does not consider the maximization of the energy efficiency, but rather analyzes the problem of minimizing the transmit power subject to rate constraints. Instead, the maximization of the fractional, bit/Joule energy efficiency is considered in [31], for the underlay scenario. There, a MIMO OFDMA system is considered, and energy-efficient precoding for the secondary system is proposed, under the simplifying assumption that different users can not employ the same transmit sub-carrier. This turns the multi-user system into a noise-limited system in which multi-user interference is completely suppressed, which allows one to solve the resource allocation problem by direct use of fractional programming.

Apart from the underlay/overlay scenario, energy efficiency in multi-user or cooperative MIMO systems has been addressed in [32] for downlink channels, and in [33], which proposes an interference neutralization technique for MIMO two-hop interference networks. In [34], full-duplex MIMO systems are considered and an iterative algorithm is proposed to maximize the energy efficiency. In [35] multiple-antenna systems are considered and the problem of network energy efficiency optimization is tackled in the asymptotic scenario of large base station antennas. In [36], a new energy-efficient performance metric is proposed, considering channel estimates at the transmitter and receiver. The relation between energy efficiency and spectral efficiency is investigated in [37]. In [38] distributed energy-efficient resource allocation is proposed for MIMO interference networks, while in [39] rate constraints are included in the energy-efficient resource allocation process for interference networks. In [40] a model for the energy efficiency of massive MIMO systems is provided, observing that a MIMO system dissipates an amount of hardware energy proportional to the number of deployed antennas.

*B. Novelty and contributions*

Motivated by the described background, this work considers both underlay and overlay communications, studying the coexistence of a multiple-input single-output (MISO) primary link with a MIMO secondary link, and tackling the problem of energy-efficient resource allocation. More in detail, the following contributions are made:

1) In the underlay operation, the problem is formulated as the maximization of the secondary link's energy efficiency, subject to a minimum rate requirement for the primary link. A maximum power constraint and a minimum rate constraint for the secondary system are also enforced. The resulting resource allocation problem is a fractional problem subject to non-convex constraints. This prevents the direct use of standard convex optimization and fractional programming techniques, and indeed no low-complexity solution method is available in the literature for this class of problems. However, we are able to determine the globally optimal solution of the problem with polynomial-time complexity.
2) The above problem is tackled assuming that rate splitting is employed at the secondary transmitter, coupled with successive interference cancellation at the secondary receiver when feasible. Although being a well-established technique to increase the achievable rate of a communication link [41], [42], the use of rate splitting in underlay communications is relatively new, and limited only to rate maximization problems [43]. The extension of this technique to energy-efficient resource allocation represents a further novelty of this work.
3) In the overlay scenario, the problem is again formulated as the maximization of the secondary link's energy efficiency, subject to a minimum rate requirement for the primary link, together with a maximum power constraint and a minimum rate constraint for the secondary system. The resulting fractional problem turns out to be more challenging than in the underlay scenario, having both non-convex constraints and an objective function which is neither concave nor pseudo-concave. Nevertheless, we provide a provably convergent algorithm which is first-order optimal and has affordable complexity. In addition, we provide a necessary and sufficient condition for the feasibility of the primary rate requirement.

4) In the overlay scenario we also provide a second algorithm, with weaker optimality claims, but with lower computational complexity. All algorithms are numerically compared, both in the underlay and overlay setting.

The rest of the paper is organized as follows. The underlay and overlay system models are described in Section II. Sections III and IV formulate the resource allocation problems and describe the proposed solutions, in the underlay and overlay scenarios, respectively. The performance of the developed algorithms is numerically addressed in Section V, while concluding remarks are given in Section VI. A background on fractional programming is provided in Appendix A.

*Notation:* Column vectors and matrices are represented in lowercase and uppercase boldface letters as $\boldsymbol{a}$ and $\boldsymbol{A}$, respectively. $\boldsymbol{A} \succeq \boldsymbol{0}$ ($\boldsymbol{A} \succ \boldsymbol{0}$) means $\boldsymbol{A}$ is positive semidefinite (positive definite). $\boldsymbol{A}^H$ is the conjugate transpose matrix of $\boldsymbol{A}$. $\boldsymbol{I}$ is the identity matrix. $\mathbb{R}$ is the set of real numbers. $|\cdot|$ is the absolute value of a scalar or the determinant of a matrix. $\text{tr}(\cdot)$ and $(\cdot)^{-1}$ stand for the trace and inverse of a matrix, respectively. $\boldsymbol{0}$ denotes a matrix whose entries are all zero. $\boldsymbol{A}^{1/2}$ denotes the unique square root of the positive semi-definite matrix $\boldsymbol{A}$.

## II. SYSTEM MODEL

Let us consider an interference channel in which a MISO primary link coexists with a MIMO secondary link. The primary system operates on a dedicated resource block, thus suffering no interference from other licensed users, and is equipped with $N_{T,1}$ antennas, while the secondary transmitter and receiver are equipped with $N_{T,2}$ and $N_R$ antennas, respectively. Denote by $\boldsymbol{h}_{1,1}$ and $\boldsymbol{H}_{2,2}$ the $N_{T,1} \times 1$ and $N_R \times N_{T,2}$ complex channels of the primary and secondary links, respectively, and by $\boldsymbol{H}_{1,2}$ and $\boldsymbol{h}_{2,1}$ the $N_R \times N_{T,1}$ channel between the primary transmitter and the secondary receiver and the $N_{T,2} \times 1$ channel between the secondary transmitter and primary receiver. All channels matrices/vectors are modeled as the product between a term $L(d, \eta)$ which accounts for large-scale effects like path-loss and shadowing, with $d$ and $\eta$ being the link distance and the power path-loss factor, and a term $\boldsymbol{\Gamma}$ which accounts for small-scale fading effects, modeled according to a quasi-static block flat fading model, possibly also including the presence of a line-of-sight component.

Thus, the particular channel realizations are transparent to the algorithms to be developed in the sequel, and the considered model is general enough to apply to many instances of communication systems in which a primary and a secondary link interfere with one another. In particular, two relevant examples are:

- *D2D cellular communications*, in which a pair of neighboring terminals bypasses the base station directly communicating with each other over a resource block used by a cellular user. In this scenario, the primary link is identified with the link between the cellular user and the base station, while the secondary link is represented by the D2D pair. The cellular user suffers no interference from other cellular users, since it uses a dedicated resource block, which is the case in present 4G systems (e.g. LTE-A networks). Similarly, the device user suffers interference only from the primary user.[1]
- *Cognitive radio systems*, in which an unlicensed user uses a frequency band licensed to a primary user, who is the legitimate owner of the spectrum. A formally equivalent interference model arises in this scenario.

**Remark 1.** *The coexistence of a MISO primary link with a MIMO secondary link is practically relevant. As for the cognitive radio setup, the primary link can model a downlink channel of a cellular system, which is typically MISO, while the secondary system is typically an ad-hoc system with different specifications than the primary system, and which could therefore be a MIMO link. As for the D2D setup, the primary link can again model a downlink channel of the legacy 3G system, while the MIMO secondary link models the channel between two 4G terminals using the LTE standard, where the use of multiple antennas is explicitly mentioned. In addition, the considered setup also applies to a MIMO primary link, with single-stream transmission and linear reception.*

For both cognitive systems and D2D scenarios, two techniques to manage the coexistence among the primary and secondary system are the underlay and overlay approach.

In the underlay protocol, the primary system is oblivious to the presence of the secondary link and wishes to support a communication rate $R_1^\star$, that must be guaranteed to its subscribers. Accordingly, the primary system tolerates a reduction of its achievable rate, provided it does not drop below the desired communication rate $R_1^\star$. In this context, a secondary system may reuse the primary resource block, provided it designs its resources in order to guarantee that the primary rate remains above the threshold $R_1^\star$ fixed by the primary system. The underlay protocol was originally introduced for cognitive radio systems [6], [44], [45], but recently found application also in D2D systems operating in the so-called reuse mode[2] [9]. In this D2D mode, a pair of neighboring devices can autonomously establish a D2D connection reusing a resource block currently used by a cellular user, provided the cellular link manages to maintain its desired communication rate.

---

[1] The scenario would be different if multiple device-to-device communications were established over the same dedicated cellular resource block. Nevertheless, it should also be stressed that one MIMO secondary link can be regarded as multiple cooperating single-antenna or single-input multiple-output (SIMO) links.

[2] It should be mentioned that the underlay approach has been also considered for other D2D operation modes involving fully centralized designs of both the cellular and device parts of the networks [46], [47]. However, these schemes do not constitute the focus of this work and will not be considered.

The overlay approach considers a different situation. An underlying assumption in the underlay approach is that the capacity of the primary communication channel (without the presence of the secondary system) is larger than the desired communication rate $R_1^\star$. In this situation, the secondary system can design its radio resources so that its interference does not cause the primary achievable rate to fall below $R_1^\star$. However, if the primary system experiences poor propagation conditions, such as high path-loss or shadowing, or if it wishes to save energy transmitting at a lower power, it might happen that even without any interference from the secondary system, the capacity of the primary channel is already below the desired rate $R_1^\star$. A way out of this problem is for the primary system to actively seek the help of a secondary transmitter, which is possibly located in a more favorable position, thus experiencing better propagation conditions.[3] In these circumstances, the secondary system becomes a relay for the primary message, and in exchange can reuse the primary resource block to transmit its own message.

Thus, the underlay and overlay protocols are to be used in complementary scenarios. If the capacity of the primary channel is large enough to allow the primary system to maintain the desired communication rate, then the underlay protocol should be used, because it is simpler, requiring no coordination between the primary and secondary systems. If instead, the point-to-point capacity of the primary channel is below the desired communication rate, then the underlay approach can not be used and the overlay protocol can be considered to increase the primary system achievable rate. In may practical scenarios, the resulting rate increase is large enough to enable the primary system to meet the desired communication rate [10], [48].

For both the underlay and overlay protocols, the primary link employs maximum ratio transmission (MRT) to maximize the primary channel point-to-point capacity, which yields:

$$R_1 = \log_2\left(1 + \frac{P_1\|\boldsymbol{h}_{1,1}\|^2}{\sigma^2}\right), \quad (1)$$

with $P_1$ and $\sigma^2$ being the primary user's transmit power and noise power at the primary receiver, modeled as $\sigma^2 = \mathcal{N}_0 BF + I_{\text{out}}$ with the first term being the thermal noise power, given by the product of the noise power spectral density $\mathcal{N}_0$, times the communication bandwidth $B$, times the receiver noise figure $F$, while the second term models the possible presence of additional interference coming from neighboring cells or systems employing the same frequency band as the system under analysis [49]. This additional interference can not be controlled since it comes from outside the system, and for this reason it is treated as noise.

### A. Underlay operation

In the underlay protocol, the received signals at the primary and secondary receivers at a given time instant are

$$y_1 = \boldsymbol{h}_{1,1}^H \boldsymbol{x}_1 + \boldsymbol{h}_{2,1}^H \boldsymbol{x}_2 + n_1, \quad (2)$$

[3]For example, some obstacles might be present between the primary transmitter and primary receiver, whereas instead the secondary transmitter might experience a better channel to the primary receiver.

$$\boldsymbol{y}_2 = \boldsymbol{H}_{2,2} \boldsymbol{x}_2 + \boldsymbol{H}_{1,2} \boldsymbol{x}_1 + \boldsymbol{n}_2, \quad (3)$$

respectively, wherein $\boldsymbol{x}_1$ and $\boldsymbol{x}_2$ are the $N_{T,1} \times 1$ and $N_{T,2} \times 1$ signal vectors sent by the primary and secondary transmitters, respectively, while $n_1 \sim \mathcal{CN}(0, \sigma^2)$ and $\boldsymbol{n}_2 \sim \mathcal{CN}(0, \sigma^2 \boldsymbol{I}_{N_R})$ model the thermal noises at the primary and secondary receiver, respectively. According to the MRT strategy we have

$$\boldsymbol{x}_1 = \sqrt{P_1} \frac{\boldsymbol{h}_{1,1}}{\|\boldsymbol{h}_{1,1}\|} s_1, \quad (4)$$

with $s_1$ the primary user's unit-modulus information symbol. As for the secondary system, the optimal receive strategy would be to use successive interference cancellation (SIC), decoding and subtracting the interference from the primary system, and then decoding the secondary message interference-free. This receive structure is known to be capacity achieving [50], and it is also optimal as far as bit/Joule energy efficiency is concerned, since it maximizes the numerator without increasing the denominator. However, for this approach to be feasible, the primary communication rate $R_1^\star$ must be achievable at the secondary receiver. If this does not hold, the standard approach would be to decode the complete secondary message treating the interference from the primary system as noise. However, it will be shown that there exist a range of $R_1^\star$ where, although conventional SIC is not not feasible, it is still possible to use SIC together with the rate-splitting approach [41], [51], which yields better performance than treating the primary message as noise. The approach is to split the secondary message into two statistically independent parts $\boldsymbol{s}_{2,1}$ and $\boldsymbol{s}_{2,2}$, which have unit-norm and are precoded by possibly different matrices $\boldsymbol{K}_{2,1}^{1/2}$ and $\boldsymbol{K}_{2,2}^{1/2}$, respectively. So, $\boldsymbol{x}_2 = \boldsymbol{K}_{2,1}^{1/2} \boldsymbol{s}_{2,1} + \boldsymbol{K}_{2,2}^{1/2} \boldsymbol{s}_{2,2}$, and even if the complete secondary message can not be decoded interference-free, it may be possible to decode interference-free at least one part of the secondary message. Specifically, the secondary receiver first decodes $\boldsymbol{s}_{2,1}$, in the presence of the interference from the primary system, then decodes and cancels the interference due to the primary signal, and finally performs an interference-free detection of $\boldsymbol{s}_{2,2}$. The analysis in Section III-B shows that there exists a range of $R_1^\star$ in which SIC is not possible for the whole secondary message, but it is feasible only for $\boldsymbol{s}_{2,2}$. Section III-B will also show that for lower values of $R_1^\star$ SIC is feasible for the complete secondary message, whereas for higher values of $R_1^\star$, not even $\boldsymbol{s}_{2,2}$ can be decoded interference-free, and the complete secondary message must be decoded in the presence of the interference from the primary system. These two cases can be regarded as extreme cases of the rate splitting approach. In the former, $\boldsymbol{s}_{2,1} = \boldsymbol{0}$ and $\boldsymbol{s}_2 = \boldsymbol{s}_{2,2}$, and SIC can be used without resorting to the rate splitting approach. In the latter, $\boldsymbol{s}_{2,2} = \boldsymbol{0}$ and $\boldsymbol{s}_2 = \boldsymbol{s}_{2,1}$. In this case SIC can not be used. Instead, linear receivers should be used, and LMMSE is considered here, as the optimal linear receiver.

### B. Overlay operation

In an overlay approach, the secondary transmitter relays the primary message in exchange for the use of the primary spectrum. As for the particular relaying protocol, any of the techniques typically employed in relay-based systems can be used, depending on the information the secondary system has about the primary system. Here, we will consider half-duplex amplify-and-forward (AF) relaying for its low-complexity, low-delay, and low-feedback requirement. Moreover, time division duplex mode is used to separate the incoming signal from the forwarded one[4]. In the first phase, the primary user employs MRT to transmit its signal $x_1$, as in the underlay scenario. During this phase, the signal $x_1$ is also received by the secondary transmitter, which, in the second phase precodes it by the $N_{T,2} \times N_{T,2}$ AF matrix $A$ and then transmits this amplified signal together with its own message $s_2$, precoded by the matrix $B^{1/2}$. The superposition of these two signals is labeled $x_2$. In this work we assume that the primary transmitter does not transmit during the second phase. It should be observed that in principle higher primary rates could be obtained if the primary user transmits in both phases. However, this would make the resource allocation problem intractable in the considered MIMO setting, requiring a prohibitive computational complexity to obtain the optimal resource allocation policy. A similar observation holds for the even more challenging approach to optimize the fraction of time during which the primary system is allowed to transmit.[5] The received signals at the primary receiver and at the secondary transmitter in the first phase are expressed as

$$y_1^{(1)} = h_{1,1}^H x_1 + n_1^{(1)}, \tag{5}$$
$$y_{st} = H_t x_1 + n_{st}, \tag{6}$$

wherein $H_t$ is the $N_{T,2} \times N_{T,1}$ channel matrix between the two transmitters, $n_1^{(1)} \sim \mathcal{CN}(0, \sigma^2)$ and $n_{st} \sim \mathcal{CN}(0, \sigma^2 I_{N_{T,2}})$ model the noises at the primary receiver and secondary transmitter, respectively, and $x_1$ is expressed as in (4). Similarly, the received signals at the primary and secondary receivers in the second phase are expressed as

$$y_1^{(2)} = h_{2,1}^H x_2 + n_1^{(2)}, \tag{7}$$
$$y_2^{(2)} = H_{2,2} x_2 + n_2, \tag{8}$$

with $n_1^{(2)} \sim \mathcal{CN}(0, \sigma^2)$ and $n_2 \sim \mathcal{CN}(0, \sigma^2 I_{N_R})$ the thermal noises at the primary and secondary receiver, respectively, and $x_2$ is given by $x_2 = A y_{st} + B^{1/2} s_2$. At the primary receiver, the data received in the two phases is stacked into the $2 \times 1$ vector

$$y_1 = \sqrt{P_1} \tilde{h}_1 s_1 + w_1 \tag{9}$$
$$= \sqrt{P_1} \begin{bmatrix} \|h_{1,1}\| \\ \frac{h_{2,1}^H A H_t h_{1,1}}{\|h_{1,1}\|} \end{bmatrix} s_1 + \begin{bmatrix} n_1^{(1)} \\ h_{2,1}^H \left( A n_{st} + B^{1/2} s_2 \right) + n_1^{(2)} \end{bmatrix},$$

which is jointly processed by LMMSE reception. We remark that since the primary system is typically not designed to cope with the additional interference from the secondary system, the use of non-linear receivers does not appear well-justified. Then, the choice of the LMMSE receiver lies in its optimality among linear receivers.

The signal received at the secondary receiver is

$$y_2^{(2)} = H_{22} \left( \frac{\sqrt{P_1}}{\|h_{1,1}\|} A H_t h_{1,1} s_1 + A n_{st} + B^{1/2} s_2 \right) + n_2. \tag{10}$$

---

[4] Since the primary and secondary link operate in the same frequency band, frequency division duplex is not feasible in the considered system.

[5] The optimal time split would inevitably depend on the optimal resource allocation at the secondary transmitter, thus requiring to alternatively optimize the time split and the secondary system radio resources.

**Remark 2.** *We stress that the relay processing is performed in the digital domain, on the base-band equivalent of the received signal [52], [53]. Unlike decode-and-forward (DF), the use of AF does not require any decoding operation at the secondary transmitter, which results in a lower complexity, a lower latency, and a simpler design. Moreover, the considered digital implementation of AF has been explicitly proposed for the LTE/A standard [54].*

In Sections III and IV, we will consider the underlay and overlay scenarios, respectively, developing resource allocation schemes for maximizing the bit/Joule energy efficiency of the secondary system, defined as the ratio between the secondary user's rate and consumed power. The optimization will be carried out subject to maximum power constraints and QoS constraints for both the primary and secondary system.

**Remark 3.** *Both in the underlay and overlay protocols that have been described in this section, the primary and secondary links are to be regarded as separate systems. In the underlay protocol, the primary and secondary system do not interact and the primary system is typically even oblivious to the presence of the secondary link, while the secondary link opportunistically reuses the primary spectrum. In the overlay protocol, the two systems cooperate to some degree, but only to achieve mutual benefit. Thus, in both cases a joint design of the primary and secondary system to optimize a common and global performance function is not possible.*

### III. RESOURCE ALLOCATION IN UNDERLAY SYSTEMS

Assume the system operates in underlay mode, as described in Section II-A. In the following we will first formulate the considered resource allocation problem, and then provide the corresponding resource allocation algorithm.

#### A. Problem formulation

In the underlay case, the resources to optimize are the secondary user's transmit covariance matrices $K_{2,1}$ and $K_{2,2}$. Let $Q_1 = \frac{P_1}{\|h_{1,1}\|^2} H_{1,2} h_{1,1} h_{1,1}^H H_{1,2}^H$. Then, the primary and secondary achievable rates are respectively expressed as

$$R_1 = \log_2 \left( 1 + \frac{P_1 \|h_{1,1}\|^2}{\sigma^2 + h_{2,1}^H (K_{2,1} + K_{2,2}) h_{2,1}} \right), \tag{11}$$

$$R_2 = \log_2 \left| I + \frac{1}{\sigma^2} H_{2,2} K_{2,2} H_{2,2}^H \right|$$
$$+ \log_2 \frac{|\sigma^2 I + H_{2,2} (K_{2,1} + K_{2,2}) H_{2,2}^H + Q_1|}{|\sigma^2 I + H_{2,2} K_{2,2} H_{2,2}^H + Q_1|}$$
$$= \log_2 \left| I + \frac{1}{\sigma^2} H_{2,2} (K_{2,1} + K_{2,2}) H_{2,2}^H + \frac{1}{\sigma^2} Q_1 \right| - R_{1,2}(K_{2,2}), \tag{12}$$

wherein $R_{1,2}$ is the primary user's achievable rate at the secondary receiver, i.e.

$$R_{1,2}(K_{2,2}) = \log_2 \left| I + \left( \sigma^2 I + H_{2,2} K_{2,2} H_{2,2}^H \right)^{-1} Q_1 \right|. \tag{13}$$

It should be mentioned that (12) also holds if rate splitting is not feasible. This case is obtained from (12) by setting $K_{2,2} = 0$.

The energy efficiency of a communication system is defined as the system benefit-cost ratio in terms of bits reliably transmitted per unit of time and Joule of consumed energy. This leads to defining the energy efficiency as the ratio between the system achievable rate and consumed power [3], namely

$$\text{EE}_2(\boldsymbol{K}_{2,1}, \boldsymbol{K}_{2,2}) = \frac{R_2(\boldsymbol{K}_{2,1}, \boldsymbol{K}_{2,2})}{\alpha \text{tr}(\boldsymbol{K}_{2,1} + \boldsymbol{K}_{2,2}) + P_c}, \quad (14)$$

with $\alpha \geq 1$ accounting for the non-ideality of the secondary user's power amplifier, and $P_c$ representing the hardware power dissipated in all other hardware components of the secondary communication system. Thus, the denominator of (14) represents the power to be consumed to ensure an achievable rate equal to the numerator. Then, the considered resource allocation problem is formulated as the optimization program

$$\max_{\boldsymbol{K}_{2,1} \succeq \boldsymbol{0}, \boldsymbol{K}_{2,2} \succeq \boldsymbol{0}} \text{EE}_2(\boldsymbol{K}_{2,1}, \boldsymbol{K}_{2,2}) \quad (15a)$$

$$\text{s.t.} \quad R_1(\boldsymbol{K}_{2,1}, \boldsymbol{K}_{2,2}) \geq R_1^\star, \quad (15b)$$

$$R_{1,2}(\boldsymbol{K}_{2,2}) \geq R_1^\star \quad (15c)$$

$$R_2(\boldsymbol{K}_{2,1}, \boldsymbol{K}_{2,2}) \geq R_2^\star \quad (15d)$$

$$\alpha \text{tr}(\boldsymbol{K}_{2,1} + \boldsymbol{K}_{2,2}) \leq P_2, \quad (15e)$$

wherein (15b) and (15d) guarantee that the primary and secondary users' rates remain above the thresholds $R_1^\star$ and $R_2^\star$, (15e) is the secondary user's power constraint, with $P_2$ the secondary user's maximum feasible power. As for (15c), the following remark is in order.

**Remark 4.** *Recalling the discussion in Section II-A, in order for SIC to be feasible at the secondary receiver (either for the complete secondary message or only for $\boldsymbol{s}_{2,2}$), the primary communication rate must be achievable at the secondary receiver. This means that the achievable rate $R_{1,2}$ of the channel between the primary transmitter and the secondary receiver must be larger than $R_1^\star$. Thus, (15c) ensures the feasibility of SIC at the secondary receiver.*

Observing that $R_{1,2}$ is convex and matrix-decreasing in $\boldsymbol{K}_{2,2}$ [55, Lemma II.3] [56, Ch 16, E.3.b.], we obtain that the objective of Problem (15) is a fractional function with concave numerator and affine denominator, and that (15d) is a concave constraint. Moreover, the power constraint in (15e) is affine, while (15b) can be reformulated into the affine constraint[6] $\boldsymbol{h}_{2,1}^H \boldsymbol{K}_{2,1} \boldsymbol{h}_{2,1} \leq P_{\text{int}}$, with

$$P_{\text{int}} = \frac{P_1 \|\boldsymbol{h}_{1,1}\|^2}{2^{R_1^\star} - 1} - \sigma^2. \quad (16)$$

The challenge in solving (15) lies in (15c), which is not a convex constraint, since, as already mentioned, $R_{1,2}$ is not a concave function of $\boldsymbol{K}_{2,2}$. As recalled in Appendix A, fractional functions can be maximized with affordable complexity provided the numerator is concave, the denominator is convex, and the constraints are convex. Therefore, directly applying standard fractional techniques to optimize (15) would result in a prohibitive computational complexity. However, in the following we show that (15) can be globally solved with polynomial complexity. Before showing this result, we make the following remarks.

**Remark 5.** *From the discussion in Section II-A it follows that in the underlay scenario $R_1^\star$ can not be larger than (1). Regarding this point, it should be explicitly observed that even if the primary user requests a rate equal to (1), i.e. $R_1^\star = \log_2(1 + \frac{P_1 \|\boldsymbol{h}_{1,1}\|^2}{\sigma^2})$, it is still possible for the secondary user to transmit a non-zero power while at the same time fulfilling the primary rate requirement, by selecting $\boldsymbol{K}_{2,1}$ and $\boldsymbol{K}_{2,2}$ such that the null-space of $\boldsymbol{K}_{2,1} + \boldsymbol{K}_{2,2}$ contains $\boldsymbol{h}_{2,1}$. Moreover, the null-space of $\boldsymbol{K}_{2,1} + \boldsymbol{K}_{2,2}$ does not change by scaling $\boldsymbol{K}_{2,1} + \boldsymbol{K}_{2,2}$ by a positive factor, thus implying that it is always possible to find $\boldsymbol{K}_{2,1}$ and $\boldsymbol{K}_{2,2}$ which fulfill the power constraint (15e), and such that the null-space of $\boldsymbol{K}_{2,1} + \boldsymbol{K}_{2,2}$ contains $\boldsymbol{h}_{2,1}$.*

**Remark 6.** *The fact that $R_1^\star$ is not larger than (1) shows that (16) is always non-negative, thereby implying that (15b) is always feasible. Instead, (15c) and (15d) might be unfeasible. However, the feasibility of (15d) might be checked through standard convex feasibility tests.[7] Instead, this is not possible for the non-convex constraint (15c), whose impact on the solution of (15) is analyzed in the rest of this section.*

### B. Resource allocation algorithm

Problem (15) will be tackled by considering three ranges of $R_1^\star$, which correspond to the extreme cases $\boldsymbol{K}_{2,2} = \boldsymbol{0}$, $\boldsymbol{K}_{2,1} = \boldsymbol{0}$, and to the intermediate case $\boldsymbol{K}_{2,2} \neq \boldsymbol{0}$, $\boldsymbol{K}_{2,1} \neq \boldsymbol{0}$.

*1) Case 1, $\boldsymbol{K}_{2,2} = \boldsymbol{0}$:* Observing that $R_{1,2}(\boldsymbol{K}_{2,2})$ is a matrix-decreasing function of $\boldsymbol{K}_{2,2}$, we obtain that $R_{1,2}$ is maximum for $\boldsymbol{K}_{2,2} = \boldsymbol{0}$, among all feasible $\boldsymbol{K}_{2,2}$. So, recalling Remark 4 it follows that (15c) is not feasible if $R_1^\star > R_{1,2}(\boldsymbol{0}) = \log_2 \left(1 + \frac{P_1 \|\boldsymbol{H}_{1,2} \boldsymbol{h}_{1,1}\|^2}{\sigma^2 \|\boldsymbol{h}_{1,1}\|^2}\right)$. This corresponds to the extreme case in which SIC is not possible even for only $\boldsymbol{s}_{2,2}$, and the complete secondary message must be decoded in the presence of the primary signal interference. Thus, in this case, the complete secondary message is encoded in $\boldsymbol{s}_{2,1}$, whereas $\boldsymbol{s}_{2,2} = \boldsymbol{0}$. This means setting $\boldsymbol{K}_{2,2} = \boldsymbol{0}$ in (15) and relaxing the unfeasible constraint (15c). Upon doing this, the optimal $\boldsymbol{K}_{2,1}$ is the solution of the fractional problem

$$\max_{\boldsymbol{K}_{2,1} \succeq \boldsymbol{0}} \frac{\log_2 \left| \boldsymbol{I} + \boldsymbol{H}_{2,2} \boldsymbol{K}_{2,1} \boldsymbol{H}_{2,2}^H \left(\sigma^2 \boldsymbol{I} + \boldsymbol{Q}_1\right)^{-1}\right|}{\alpha \text{tr}(\boldsymbol{K}_{2,1}) + P_c} \quad (17a)$$

$$\text{s.t.} \quad \boldsymbol{h}_{2,1}^H \boldsymbol{K}_{2,1} \boldsymbol{h}_{2,1} \leq P_{\text{int}}, \quad (17b)$$

$$\alpha \text{tr}(\boldsymbol{K}_{2,1}) \leq P_2 \quad (17c)$$

$$\log_2 \left| \boldsymbol{I} + \boldsymbol{H}_{2,2} \boldsymbol{K}_{2,1} \boldsymbol{H}_{2,2}^H \left(\sigma^2 \boldsymbol{I} + \boldsymbol{Q}_1\right)^{-1}\right| \geq R_2^\star \quad (17d)$$

Problem (17) has affine constraints, while the objective has a concave numerator and an affine denominator. As a consequence, (17) can be globally solved with polynomial complexity by means of fractional programming theory.

*2) Case 2, $\boldsymbol{K}_{2,1} = \boldsymbol{0}$:* This scenario corresponds to the extreme case in which SIC at the secondary receiver is feasible for the complete secondary message. So, the complete secondary message can be encoded into $\boldsymbol{s}_{2,2}$, while $\boldsymbol{s}_{2,1} = \boldsymbol{0}$, and thus $\boldsymbol{K}_{2,1} = \boldsymbol{0}$ in Problem (15). Proposition 1 shows that this extreme case corresponds to the following range of $R_1^\star$:

$$R_1^\star \leq \log_2 \left| \boldsymbol{I} + \boldsymbol{Q}_1 (\sigma^2 \boldsymbol{I} + \boldsymbol{H}_{2,2} \boldsymbol{\Sigma}^\star \boldsymbol{H}_{2,2}^H)^{-1} \right|, \quad (18)$$

---

[6]Note that this would not be possible in case of a MIMO primary link.

[7]If (15d) is not feasible, the secondary user must accept a lower $R_2^\star$. In the following we assume that the chosen $R_2^\star$ results in a feasible (15d).

with $\boldsymbol{\Sigma}^\star$ the solution to the problem

$$\max_{\boldsymbol{K}_{2,2} \succeq \boldsymbol{0}} \frac{\log_2 \left|\boldsymbol{I} + \frac{1}{\sigma^2}\boldsymbol{H}_{2,2}\boldsymbol{K}_{2,2}\boldsymbol{H}_{2,2}^H\right|}{\alpha \mathrm{tr}(\boldsymbol{K}_{2,2}) + P_c} \quad (19\mathrm{a})$$

$$\text{s.t.} \quad \boldsymbol{h}_{2,1}^H \boldsymbol{K}_{2,2} \boldsymbol{h}_{2,1} \leq P_{\mathrm{int}}, \quad (19\mathrm{b})$$

$$\alpha \mathrm{tr}(\boldsymbol{K}_{2,2}) \leq P_2 \quad (19\mathrm{c})$$

$$\log_2 \left|\boldsymbol{I} + \frac{1}{\sigma^2}\boldsymbol{H}_{2,2}\boldsymbol{K}_{2,2}\boldsymbol{H}_{2,2}^H\right| \geq R_2^\star \quad (19\mathrm{d})$$

and $P_{\mathrm{int}}$ as defined in (16). We observe that computing (18) can be managed with affordable complexity, since Problem (19) has a convex feasibility set, while the objective has a concave numerator and an affine denominator. Thus, (19) can be globally solved with polynomial complexity by means of fractional programming theory.

**Proposition 1.** *Let $(\boldsymbol{K}_{2,1}^\star, \boldsymbol{K}_{2,2}^\star)$ be a solution of (15). If (18) holds, then $(\boldsymbol{K}_{2,1}^\star = \boldsymbol{0}, \boldsymbol{K}_{2,2}^\star = \boldsymbol{\Sigma}^\star)$.*

*Proof:* Let us first consider the following relaxed version of Problem (15) without the constraint (15c)

$$\max_{\boldsymbol{K}_{2,1} \succeq \boldsymbol{0}, \boldsymbol{K}_{2,2} \succeq \boldsymbol{0}} \frac{R_2(\boldsymbol{K}_{2,1}, \boldsymbol{K}_{2,2})}{\alpha \mathrm{tr}(\boldsymbol{K}_{2,1} + \boldsymbol{K}_{2,2}) + P_c} \quad (20\mathrm{a})$$

$$\text{s.t.} \quad \boldsymbol{h}_{2,1}^H(\boldsymbol{K}_{2,1} + \boldsymbol{K}_{2,2})\boldsymbol{h}_{2,1} \leq P_{\mathrm{int}} \quad (20\mathrm{b})$$

$$\alpha \mathrm{tr}(\boldsymbol{K}_{2,1} + \boldsymbol{K}_{2,2}) \leq P_2 \quad (20\mathrm{c})$$

$$R_2(\boldsymbol{K}_{2,1}, \boldsymbol{K}_{2,2}) \geq R_2^\star \quad (20\mathrm{d})$$

with $P_{\mathrm{int}}$ as defined in (16). Let us denote by $(\widetilde{\boldsymbol{K}}_{2,1}, \widetilde{\boldsymbol{K}}_{2,2})$ the optimal solution of (20). The first step of the proof is to show that $\widetilde{\boldsymbol{K}}_{2,1} = \boldsymbol{0}$. This is accomplished by a contradiction argument, and in particular by showing that if $\widetilde{\boldsymbol{K}}_{2,1}$ is not the zero matrix, then we can find another feasible point of (20) which yields a higher objective (20a) than $(\widetilde{\boldsymbol{K}}_{2,1}, \widetilde{\boldsymbol{K}}_{2,2})$. To show this, let us assume that $\widetilde{\boldsymbol{K}}_{2,1}$ has at least one strictly positive eigenvalue[8]. Then there exists $\boldsymbol{\Psi} \succ \boldsymbol{0}$ such that we can construct the matrix $\bar{\boldsymbol{K}}_{2,1} = \widetilde{\boldsymbol{K}}_{2,1} - \boldsymbol{\Psi} \succeq \boldsymbol{0}$. Moreover, since $\boldsymbol{\Psi} \succ \boldsymbol{0}$ we can also construct the matrix $\bar{\boldsymbol{K}}_{2,2} = \widetilde{\boldsymbol{K}}_{2,2} + \boldsymbol{\Psi} \succ \widetilde{\boldsymbol{K}}_{2,2}$. At this point we observe that the two matrices we have constructed are such that $\bar{\boldsymbol{K}}_{2,1} + \bar{\boldsymbol{K}}_{2,2} = \widetilde{\boldsymbol{K}}_{2,1} + \widetilde{\boldsymbol{K}}_{2,2}$. Thus, the pair $(\bar{\boldsymbol{K}}_{2,1}, \bar{\boldsymbol{K}}_{2,2})$ is feasible and it yields the same denominator in (20a) as $(\widetilde{\boldsymbol{K}}_{2,1}, \widetilde{\boldsymbol{K}}_{2,2})$. Moreover, given the expression of $R_2$ in (12), recalling that $R_{1,2}$ is matrix-decreasing in $\boldsymbol{K}_{2,2}$, and since $\bar{\boldsymbol{K}}_{2,2} \succ \widetilde{\boldsymbol{K}}_{2,2}$, it follows that $R_2(\bar{\boldsymbol{K}}_{1,2}, \bar{\boldsymbol{K}}_{2,2}) > R_2(\widetilde{\boldsymbol{K}}_{1,2}, \widetilde{\boldsymbol{K}}_{2,2})$. Therefore, $(\bar{\boldsymbol{K}}_{1,2}, \bar{\boldsymbol{K}}_{2,2})$ is feasible and yields a larger (20a) than $(\widetilde{\boldsymbol{K}}_{2,1}, \widetilde{\boldsymbol{K}}_{2,2})$. This is a contradiction and consequently $\widetilde{\boldsymbol{K}}_{2,1} = \boldsymbol{0}$.

Then, there is no loss of optimality in setting $\boldsymbol{K}_{2,1} = \boldsymbol{0}$ in (20). Doing this yields (19), whose optimal solution has been denoted by $\boldsymbol{\Sigma}^\star$. Finally, the pair $(\boldsymbol{K}_{2,1}^\star = \boldsymbol{0}, \boldsymbol{K}_{2,2}^\star = \boldsymbol{\Sigma}^\star)$ is also feasible for the original problem (15) because of (18), and therefore it is also the solution of the original problem (15). ∎

To conclude Case 2, we provide the following result.

**Lemma 1.** *Problem* (19) *has a unique solution.*

*Proof:* Consider the numerator of the objective function (19a). If $\boldsymbol{H}_{22}^H \boldsymbol{H}_{22} \succ \boldsymbol{0}$, then $\log_2 \left|\boldsymbol{I} + \frac{1}{\sigma^2}\boldsymbol{H}_{22}\boldsymbol{K}_{2,2}\boldsymbol{H}_{22}^H\right|$

---

[8]This is equivalent to contradicting $\widetilde{\boldsymbol{K}}_{2,1} = \boldsymbol{0}$ in the space of positive semidefinite matrices.

is strictly concave in $\boldsymbol{K}_{2,2}$ [55, Lemma II.4]. Since the denominator of (19a) is affine, this implies that (19a) is a strictly pseudo-concave function, and thus Problem (19) has a unique solution. On the other hand, if $\boldsymbol{H}_{22}^H\boldsymbol{H}_{22}$ has a null-space, no direction of $\boldsymbol{K}_{2,2}$ should lie in such null space in order to maximize (19a). Thus, without loss of optimality, it is possible consider only the directions of $\boldsymbol{H}_{22}$ which correspond to positive eigenvalues of $\boldsymbol{H}_{22}^H\boldsymbol{H}_{22}$, which implies that (19) has a unique solution. ∎

*3) Case 3, $\boldsymbol{K}_{2,1} \neq \boldsymbol{0}$ and $\boldsymbol{K}_{2,2} \neq \boldsymbol{0}$:* If both $\boldsymbol{K}_{2,1} \neq \boldsymbol{0}$ and $\boldsymbol{K}_{2,2} \neq \boldsymbol{0}$, then SIC at the secondary receiver is not possible, and only one part of the secondary message can be decoded after removing the interference due to the primary signal. Non-zero matrices $\boldsymbol{K}_{2,1}$ and $\boldsymbol{K}_{2,2}$ can be obtained when $R_1^\star$ lies in the following range.

$$\log_2\left|\boldsymbol{I}+\boldsymbol{Q}_1(\sigma^2\boldsymbol{I}+\boldsymbol{H}_{2,2}\boldsymbol{\Sigma}^\star\boldsymbol{H}_{2,2}^H)^{-1}\right| < R_1^\star \leq \log_2\left(1+\frac{P_1\|\boldsymbol{H}_{1,2}\boldsymbol{h}_{1,1}\|^2}{\sigma^2\|\boldsymbol{h}_{1,1}\|^2}\right) \quad (21)$$

In this case, the solution of (15) is given by the following result.

**Proposition 2.** *If (21) holds, then the optimal $\boldsymbol{K}_{2,1}$ and $\boldsymbol{K}_{2,2}$ for Problem (15) are given by:*

$$\boldsymbol{K}_{2,1} = \widehat{\boldsymbol{K}}_{2,1} + (1-\gamma)\widehat{\boldsymbol{K}}_{2,2} \quad (22)$$

$$\boldsymbol{K}_{2,2} = \gamma \widehat{\boldsymbol{K}}_{2,2}, \quad (23)$$

*with $\gamma \in [0,1]$ such that $R_{1,2}(\boldsymbol{K}_{2,2}) = R_1^\star$, and $\widehat{\boldsymbol{K}}_{2,1}$, $\widehat{\boldsymbol{K}}_{2,2}$ being the solution of the following problem:*

$$\max_{\boldsymbol{K}_{2,1} \succeq \boldsymbol{0}, \boldsymbol{K}_{2,2} \succeq \boldsymbol{0}} \frac{\log_2\left|\boldsymbol{I}+\frac{1}{\sigma^2}\boldsymbol{H}_{2,2}(\boldsymbol{K}_{2,1}+\boldsymbol{K}_{2,2})\boldsymbol{H}_{2,2}^H+\frac{1}{\sigma^2}\boldsymbol{Q}_1\right| - R_1^\star}{\alpha \mathrm{tr}(\boldsymbol{K}_{2,1}+\boldsymbol{K}_{2,2}) + P_c} \quad (24\mathrm{a})$$

$$\text{s.t.} \ \boldsymbol{h}_{2,1}^H(\boldsymbol{K}_{2,1}+\boldsymbol{K}_{2,2})\boldsymbol{h}_{2,1} \leq P_{\mathrm{int}}, \ \alpha \mathrm{tr}(\boldsymbol{K}_{2,1}+\boldsymbol{K}_{2,2}) \leq P_2, \quad (24\mathrm{b})$$

$$\log_2\left|\boldsymbol{I}+\boldsymbol{Q}_1(\sigma^2\boldsymbol{I}+\boldsymbol{H}_{2,2}\boldsymbol{K}_{2,2}\boldsymbol{H}_{2,2}^H)^{-1}\right| \leq R_1^\star, \quad (24\mathrm{c})$$

$$\log_2\left|\boldsymbol{I}+\frac{1}{\sigma^2}\boldsymbol{H}_{2,2}(\boldsymbol{K}_{2,1}+\boldsymbol{K}_{2,2})\boldsymbol{H}_{2,2}^H+\frac{1}{\sigma^2}\boldsymbol{Q}_1\right| - R_1^\star \geq R_2^\star \quad (24\mathrm{d})$$

*Proof:* See Appendix B. ∎

Problem (24) can be solved with polynomial complexity by means of fractional programming, since the numerator and denominator of the objective are respectively concave and affine, while the constraint set is convex.

Putting all three cases together, the overall energy-efficient resource allocation algorithm for the underlay scenario can be formally stated as in Algorithm 1. Since all three cases could be globally solved with polynomial complexity by using fractional programming, Algorithm 1 is guaranteed to yield the global solution of Problem (15), while at the same time having polynomial computational complexity.

**Remark 7.** *We stress that Algorithm 1 can be straightforwardly specialized to perform rate maximization, by setting $\alpha = 0$, $P_c = 1$ in the expression of the energy efficiency.*

**Remark 8.** *Algorithm 1 is implemented at the secondary transmitter, which is required to know $\boldsymbol{H}_{2,2}$, $\boldsymbol{h}_{2,1}$, $\boldsymbol{Q}_1$, and $P_1\|\boldsymbol{h}_{1,1}\|^2$. Standard methods can be used to estimate $\boldsymbol{H}_{2,2}$ as well as the equivalent channel $\sqrt{P_1}\boldsymbol{H}_{1,2}\boldsymbol{h}_{1,1}/\|\boldsymbol{h}_{1,1}\|$, from which $\boldsymbol{Q}_1$ can be obtained. Assuming reciprocity, $\boldsymbol{h}_{2,1}$ can be estimated from the feedback transmissions of the primary*

**Algorithm 1** $EE_2$ maximization for underlay communications.

**if** $R_1^\star \geq \log_2\left(1 + \frac{P_1\|\boldsymbol{H}_{1,2}\boldsymbol{h}_{1,1}\|^2}{\sigma^2\|\boldsymbol{h}_{1,1}\|^2}\right)$ **then**
$\quad \boldsymbol{K}_{2,2} = \boldsymbol{0}$; Set $\boldsymbol{K}_{2,1}$ as the solution of (17).
**else**
$\quad$ Set $\boldsymbol{\Sigma}^\star$ as the solution of (19);
$\quad$ **if** $R_1^\star \leq \log_2\left|\boldsymbol{I} + \boldsymbol{Q}_1(\sigma^2\boldsymbol{I} + \boldsymbol{H}_{2,2}\boldsymbol{\Sigma}^\star\boldsymbol{H}_{2,2}^H)^{-1}\right|$ **then**
$\quad\quad \boldsymbol{K}_{2,1} = \boldsymbol{0}, \quad \boldsymbol{K}_{2,2} = \boldsymbol{\Sigma}^\star$
$\quad$ **else**
$\quad\quad$ Set $(\widehat{\boldsymbol{K}}_{2,1}, \widehat{\boldsymbol{K}}_{2,2})$ as the solution of (24);
$\quad\quad$ Find $\widehat{\gamma} \in [0; 1]$ such that $R_{1,2}(\widehat{\gamma}\widehat{\boldsymbol{K}}_{2,2}) = R_1^\star$.
$\quad\quad \boldsymbol{K}_{2,1} = \widehat{\boldsymbol{K}}_{2,1} + (1-\widehat{\gamma})\widehat{\boldsymbol{K}}_{2,2}, \quad \boldsymbol{K}_{2,2} = \widehat{\gamma}\widehat{\boldsymbol{K}}_{2,2}$
$\quad$ **end if**
**end if**

*receiver. As for $P_1\|\boldsymbol{h}_{1,1}\|^2$, this is the channel quality indicator (CQI) of the primary system, which is a scalar parameter regularly fed back by the primary receiver. Then, the secondary transmitter can overhear this transmission and decode the information for the value of $P_1\|\boldsymbol{h}_{1,1}\|^2$.*

## IV. RESOURCE ALLOCATION IN OVERLAY SYSTEMS

In this section we focus on the overlay scenario. As in Section III, we will first formulate the resource allocation problem, and then describe the corresponding resource allocation algorithms.

### A. Problem formulation

In the overlay case, the resources to allocate are the secondary user's AF matrix $\boldsymbol{A}$ and precoding matrix $\boldsymbol{B}^{1/2}$. Given the system model described in Section II-B, the secondary user's achievable rate is written as

$$R_2(\boldsymbol{A}, \boldsymbol{B}) = \frac{1}{2}\log_2|\boldsymbol{I} + \boldsymbol{Z}^{-1/2}\boldsymbol{H}_{2,2}\boldsymbol{B}\boldsymbol{H}_{2,2}^H\boldsymbol{Z}^{-1/2}|, \quad (25)$$

with $\boldsymbol{Z}$ being the interference-plus-noise covariance matrix at the secondary receiver $\boldsymbol{Z}(\boldsymbol{A}) = \sigma^2\boldsymbol{I}_{N_R} + \boldsymbol{H}_{2,2}\boldsymbol{A}\boldsymbol{M}\boldsymbol{A}^H\boldsymbol{H}_{2,2}^H$, wherein $\boldsymbol{M} = \frac{P_1}{\|\boldsymbol{h}_{1,1}\|^2}\boldsymbol{H}_t\boldsymbol{h}_{1,1}\boldsymbol{h}_{1,1}^H\boldsymbol{H}_t^H + \sigma^2\boldsymbol{I}_{N_{T,2}}$. The corresponding energy efficiency (EE) of the secondary user is

$$EE_2 = \frac{R_2(\boldsymbol{A}, \boldsymbol{B})}{\alpha\text{tr}(\boldsymbol{A}\boldsymbol{M}\boldsymbol{A}^H + \boldsymbol{B}) + P_c}, \quad (26)$$

where $\text{tr}(\boldsymbol{A}\boldsymbol{M}\boldsymbol{A}^H)$ is the transmit power used by the secondary transmitter to forward the signal from the primary user.

Next, recalling (9), the primary user's interference-plus-noise covariance matrix $\boldsymbol{W}$ is

$$\boldsymbol{W} = \mathbb{E}\left[\boldsymbol{w}_1\boldsymbol{w}_1^H\right] = \begin{bmatrix} \sigma^2 & 0 \\ 0 & \boldsymbol{h}_{2,1}^H(\sigma^2\boldsymbol{A}\boldsymbol{A}^H + \boldsymbol{B})\boldsymbol{h}_{2,1} + \sigma^2 \end{bmatrix}, \quad (27)$$

and the resulting achievable rate is given by,

$$R_1 = \frac{1}{2}\log_2(1 + P_1\tilde{\boldsymbol{h}}_1^H\boldsymbol{W}^{-1}\tilde{\boldsymbol{h}}_1) \quad (28)$$
$$= \frac{1}{2}\log_2\left(1 + \frac{P_1\|\boldsymbol{h}_{1,1}\|^2}{\sigma^2} + \frac{P_1}{\|\boldsymbol{h}_{1,1}\|^2}\frac{|\boldsymbol{h}_{2,1}^H\boldsymbol{A}\boldsymbol{H}_t\boldsymbol{h}_{1,1}|^2}{\sigma^2 + \boldsymbol{h}_{2,1}^H(\sigma^2\boldsymbol{A}\boldsymbol{A}^H + \boldsymbol{B})\boldsymbol{h}_{2,1}}\right)$$

where the factor $1/2$ stems from the fact that each primary user's symbol spans two symbol intervals, due to the half-duplex operation of the secondary transmitter. Finally, the energy-efficient resource allocation problem is formulated as

$$\max_{\boldsymbol{A},\boldsymbol{B}\succeq \boldsymbol{0}} \quad EE_2 \quad (29a)$$
$$\text{s.t.} \quad R_1 \geq R_1^\star, \quad (29b)$$
$$\alpha\text{tr}(\boldsymbol{A}\boldsymbol{M}\boldsymbol{A}^H + \boldsymbol{B}) \leq P_2, \quad (29c)$$
$$R_2(\boldsymbol{K}_{2,1}, \boldsymbol{K}_{2,2}) \geq R_2^\star, \quad (29d)$$

wherein $P_2$ is the secondary user's maximum feasible transmit power and $R_1^\star$ is the minimum achievable rate that the primary user can accept. Similarly to the underlay scenario, Problem (29) is a fractional, matrix-valued, problem in which it is not possible to jointly diagonalize both the objective and the constraints due to the presence of different channels. As a result, it is not possible to simplify the problem by deriving the transmit directions of $\boldsymbol{A}$ and $\boldsymbol{B}$ in closed-form. Moreover, Problem (29) poses additional challenges with respect to its underlay counterpart (15):

**1)** Unlike (15), the secondary user's rate which appears both at the numerator of (29) and in (29d) is not jointly concave in the optimization variables, as it is not concave in $\boldsymbol{A}$.

**2)** The challenge in solving (15) was related to the non-convex constraint (15c) stemming from the use of successive interference cancellation at the secondary receiver. In the overlay scenario, (15c) is not enforced since no successive decoding is used at the secondary receiver. Nevertheless, the constraint set is still not convex, due to (29b) and (29d).

**3)** In the underlay scenario, the non-convex constraint (15c) was depending on only one of the two optimization variables, and was a decreasing function. These properties were critical to develop the solution method. Instead, (29b) depends on both optimization variables and it does not enjoy any clear monotonicity property with respect to $\boldsymbol{A}$. For all these reasons, the approach used in the underlay setting does not work in the overlay scenario, and globally solving (29) with an affordable complexity appears to be a challenging task. In the coming section IV-C, we will trade-off performance with complexity, providing two sub-optimal, low-complexity algorithms. The former enjoys Karush Kuhn Tucker (KKT) optimality properties, requiring only the solution of a sequence of easier fractional problems. The latter has weaker optimality claims but an even lower-computational complexity. Numerical results will show that the two approaches enjoy very similar performance.

### B. Feasibility of (29)

The first step towards the analysis of (29) is to study its feasibility. Problem (29) might be unfeasible due to either (29b) or (29d). A joint feasibility analysis of both constraints appears challenging. In this section the focus will be the rate guarantee for the primary user in (29b), which is more critical than the rate guarantee for the secondary user in (29d). Indeed, if (29b) is not fulfilled, the secondary user can not transmit at all. Instead, if (29d) is not fulfilled, the secondary user can still transmit, but must accept a lower rate guarantee $R_2^\star$.

Recalling the discussion from Section II-B, we see that in the overlay scenario $R_1^\star$ should be larger than (1). As a result, Problem (29) might turn out to be unfeasible if a too large $R_1^\star$ is required, and thus the feasibility range of $R_1^\star$ should be determined. To this end, for any given $P_2$, the maximum $R_1^\star$ which can be guaranteed is determined in the following proposition.

**Proposition 3.** *For any fixed $P_2$, Problem (29) is feasible provided $R_1^\star \leq \overline{R}_1 = R_1(\boldsymbol{B} = \boldsymbol{0}, \boldsymbol{A}^\star)$, wherein*

$$\boldsymbol{A}^\star = \sqrt{a}\,\frac{\boldsymbol{h}_{2,1}}{\|\boldsymbol{h}_{2,1}\|}\,\frac{\boldsymbol{h}_{1,1}^H \boldsymbol{H}_t^H}{\|\boldsymbol{H}_t \boldsymbol{h}_{1,1}\|}, \quad a = \frac{\|\boldsymbol{h}_{1,1}\|^2 P_2/\alpha}{\|\boldsymbol{H}_t \boldsymbol{h}_{1,1}\|^2 P_1 + \sigma^2 \|\boldsymbol{h}_{1,1}\|^2}\,. \tag{30}$$

*Proof:* See Appendix C. ∎

**Remark 9.** *From Proposition 3, and since $R_1^\star$ should be larger than (1), we obtain that $R_1^\star$ should lie in the interval:*

$$\log_2\left(1 + \frac{P_1 \|\boldsymbol{h}_{1,1}\|^2}{\sigma^2}\right) < R_1^\star \leq R_1(\boldsymbol{B} = \boldsymbol{0}, \boldsymbol{A}^\star)\,. \tag{31}$$

*In the context of D2D communications, (31) implies that the overlay approach has a two-fold advantage. Not only the base station has one free resource block because the D2D user is reusing the resource block of another cellular user, but the overall rate of the system increases. Therefore, an additional user can be admitted into the system and can be scheduled to the resource block left unused by the D2D user.*

### C. Resource allocation algorithms

As already mentioned, directly using fractional programming to solve (29) would result in a prohibitive complexity. For this reason, in the following we will resort to the tool of sequential fractional programming. The idea is to tackle (29) by solving a sequence of easier fractional problems. To this end, we recall the following result.

**Proposition 4** ([57])**.** *Consider an optimization problem $\mathcal{P}$ with continuously differentiable objective $f_0$ and constraint functions $\{f_i\}_{i\geq 1}$. Assume that a sequence of optimization problems $\{\mathcal{P}_\ell\}_\ell$ exists, with continuously differentiable objectives $\{f_{0,\ell}\}_\ell$ and constraint functions $\{f_{i,\ell}\}_{i,\ell}$, and with solutions $\{\boldsymbol{x}^{(\ell)}\}$. If the following three properties are fulfilled, for all $\ell$ and all $i \geq 0$:*

(**P1**) $f_{i,\ell}(\boldsymbol{x}) \leq f_i(\boldsymbol{x})$, *for all $\boldsymbol{x}$,*
(**P2**) $f_{i,\ell}(\boldsymbol{x}^{(\ell-1)}) = f_i(\boldsymbol{x}^{(\ell-1)})$,
(**P3**) $\nabla f_{i,\ell}(\boldsymbol{x}^{(\ell-1)}) = \nabla f_i(\boldsymbol{x}^{(\ell-1)})$,

*then, the sequence $\{f_0(\boldsymbol{x}^{(\ell)})\}_\ell$ is monotonically increasing and converges. Moreover, upon convergence, the objective value is equal to the value at a KKT point of Problem $\mathcal{P}$.*

Leveraging Proposition 4, we can monotonically increase the value of the energy efficiency in (29a), until we attain first-order optimality. Of course, the critical point in this approach is to find suitable lower-bounds of (29a) and (29d), which fulfill Properties (**P1**), (**P2**), and (**P3**), while at the same time leading to a simpler maximization problem. In the following, we accomplish this first for Problem (29), and then by considering a simplified version of (29).

*1) Sequential fractional programming for (29):* Before focusing on finding suitable lower-bounds for (29a) and (29d), we apply the substitution $\boldsymbol{A}\boldsymbol{M}^{1/2} = \boldsymbol{U}\boldsymbol{\Lambda}^{1/2}\boldsymbol{V}^H$ to Problem (29), with $\boldsymbol{U}\boldsymbol{\Lambda}^{1/2}\boldsymbol{\Lambda}^{H/2}\boldsymbol{U}^H = \boldsymbol{X}$, which yields[9]

$$\max_{\substack{\boldsymbol{X}\succeq \boldsymbol{0} \\ \boldsymbol{B}\succeq \boldsymbol{0}}} \frac{\log_2\left|\sigma^2 \boldsymbol{I} + \boldsymbol{H}_{2,2}(\boldsymbol{X}+\boldsymbol{B})\boldsymbol{H}_{2,2}^H\right| - \log_2\left|\sigma^2 \boldsymbol{I} + \boldsymbol{H}_{2,2}\boldsymbol{X}\boldsymbol{H}_{2,2}^H\right|}{\alpha \mathrm{tr}(\boldsymbol{X}+\boldsymbol{B}) + P_c} \tag{32a}$$

$$\text{s.t. } c^\star(\sigma^2 + \boldsymbol{h}_{2,1}^H \boldsymbol{B} \boldsymbol{h}_{2,1}) \leq \boldsymbol{h}_{2,1}^H \boldsymbol{X} \boldsymbol{h}_{2,1}$$
$$- (c^\star + 1)\sigma^2 \boldsymbol{h}_{2,1}^H \boldsymbol{U} \boldsymbol{\Lambda}^{1/2} \boldsymbol{V}^H \boldsymbol{M}^{-1} \boldsymbol{V} \boldsymbol{\Lambda}^{H/2} \boldsymbol{U}^H \boldsymbol{h}_{2,1} \tag{32b}$$

$$\alpha \mathrm{tr}(\boldsymbol{X}+\boldsymbol{B}) \leq P_2 \tag{32c}$$

$$\log_2\left|\sigma^2 \boldsymbol{I} + \boldsymbol{H}_{2,2}(\boldsymbol{X}+\boldsymbol{B})\boldsymbol{H}_{2,2}^H\right| - \log_2\left|\sigma^2 \boldsymbol{I} + \boldsymbol{H}_{2,2}\boldsymbol{X}\boldsymbol{H}_{2,2}^H\right| \geq R_2^\star \tag{32d}$$

wherein $c^\star = 2^{2R_1^\star} - 1 - \frac{P_1 \|\boldsymbol{h}_{1,1}\|^2}{\sigma^2}$. At this point, we can determine the optimal $\boldsymbol{V}$.

**Proposition 5.** *For any $\boldsymbol{U}$ and $\boldsymbol{\Lambda}$, the optimal $\boldsymbol{V}$ is such that $\boldsymbol{V}^H \boldsymbol{H}_t \boldsymbol{h}_{1,1}$ and $\boldsymbol{\Lambda}^{H/2}\boldsymbol{U}^H \boldsymbol{h}_{2,1}$ are parallel vectors, and the corresponding expression of (32b) is*

$$\frac{(c^\star+1)P_1 \|\boldsymbol{H}_t \boldsymbol{h}_{1,1}\|^2}{\sigma^2 \|\boldsymbol{h}_{1,1}\|^2 + P_1 \|\boldsymbol{H}_t \boldsymbol{h}_{1,1}\|^2} \boldsymbol{h}_{2,1}^H \boldsymbol{X} \boldsymbol{h}_{2,1} \geq c^\star(\sigma^2 + \boldsymbol{h}_{2,1}^H(\boldsymbol{B}+\boldsymbol{X})\boldsymbol{h}_{2,1})$$

*Proof:* In Problem (32), the variable $\boldsymbol{V}$ appears only in (32b), which, plugging the expression of $\boldsymbol{M}$ and using the matrix inversion lemma, can be expressed as

$$c^\star \sigma^2 + \boldsymbol{h}_{2,1}^H (c^\star \boldsymbol{B} - \boldsymbol{X}) \boldsymbol{h}_{2,1}$$
$$\leq -(c^\star + 1)\boldsymbol{h}_{2,1}^H \boldsymbol{U} \boldsymbol{\Lambda}^{1/2} \boldsymbol{V}^H$$
$$\left[\boldsymbol{I} - \frac{P_1 \boldsymbol{H}_t \boldsymbol{h}_{1,1}\boldsymbol{h}_{1,1}^H \boldsymbol{H}_t^H}{\sigma^2 \|\boldsymbol{h}_{1,1}\|^2 + P_1 \|\boldsymbol{H}_t \boldsymbol{h}_{1,1}\|^2}\right] \boldsymbol{V} \boldsymbol{\Lambda}^{H/2} \boldsymbol{U}^H \boldsymbol{h}_{2,1} \tag{33}$$

$$c^\star(\sigma^2 + \boldsymbol{h}_{2,1}^H(\boldsymbol{B}+\boldsymbol{X})\boldsymbol{h}_{2,1})$$
$$\leq \frac{(c^\star+1)P_1}{\sigma^2 \|\boldsymbol{h}_{1,1}\|^2 + P_1 \|\boldsymbol{H}_t \boldsymbol{h}_{1,1}\|^2} \left|\boldsymbol{h}_{2,1}^H \boldsymbol{U} \boldsymbol{\Lambda}^{1/2} \boldsymbol{V}^H \boldsymbol{H}_t \boldsymbol{h}_{1,1}\right|^2 \tag{34}$$

At this point, since $\boldsymbol{V}$ does not depend on $\boldsymbol{X}$, the optimal $\boldsymbol{V}$ is such that the right-hand side (RHS) of (33) is maximized. By Cauchy-Schwarz inequality, this happens by choosing the rotation matrix $\boldsymbol{V}$ in order to make $\boldsymbol{V}^H \boldsymbol{H}_t \boldsymbol{h}_{1,1}$ parallel to $\boldsymbol{\Lambda}^{H/2}\boldsymbol{U}^H \boldsymbol{h}_{2,1}$. For this choice of $\boldsymbol{V}$, we have

$$|\boldsymbol{h}_{2,1}^H \boldsymbol{U} \boldsymbol{\Lambda}^{1/2} \boldsymbol{V}^H \boldsymbol{H}_t \boldsymbol{h}_{1,1}|^2 = \|\boldsymbol{h}_{2,1}^H \boldsymbol{U} \boldsymbol{\Lambda}^{1/2}\|^2 \|\boldsymbol{H}_t \boldsymbol{h}_{1,1}\|^2$$
$$= \boldsymbol{h}_{2,1}^H \boldsymbol{X} \boldsymbol{h}_{2,1} \|\boldsymbol{H}_t \boldsymbol{h}_{1,1}\|^2, \tag{35}$$

and plugging (35) into (33) we obtain the thesis. ∎

As a result of Proposition 5, there is no loss of optimality in recasting (32) as

$$\max_{\substack{\boldsymbol{X}\succeq \boldsymbol{0} \\ \boldsymbol{B}\succeq \boldsymbol{0}}} \frac{\log_2\left|\sigma^2 \boldsymbol{I} + \boldsymbol{H}_{2,2}(\boldsymbol{X}+\boldsymbol{B})\boldsymbol{H}_{2,2}^H\right| - \log_2\left|\sigma^2 \boldsymbol{I} + \boldsymbol{H}_{2,2}\boldsymbol{X}\boldsymbol{H}_{2,2}^H\right|}{\alpha \mathrm{tr}(\boldsymbol{X}+\boldsymbol{B}) + P_c} \tag{36a}$$

$$\text{s.t. } \frac{(c^\star+1)P_1\|\boldsymbol{H}_t\boldsymbol{h}_{1,1}\|^2 \boldsymbol{h}_{2,1}^H \boldsymbol{X} \boldsymbol{h}_{2,1}}{c^\star \sigma^2 \|\boldsymbol{h}_{1,1}\|^2 + c^\star P_1 \|\boldsymbol{H}_t \boldsymbol{h}_{1,1}\|^2} \geq \sigma^2 + \boldsymbol{h}_{2,1}^H(\boldsymbol{B}+\boldsymbol{X})\boldsymbol{h}_{2,1} \tag{36b}$$

$$\alpha \mathrm{tr}(\boldsymbol{X}+\boldsymbol{B}) \leq P_2 \tag{36c}$$

$$\log_2\left|\sigma^2 \boldsymbol{I} + \boldsymbol{H}_{2,2}(\boldsymbol{X}+\boldsymbol{B})\boldsymbol{H}_{2,2}^H\right| - \log_2\left|\sigma^2 \boldsymbol{I} + \boldsymbol{H}_{2,2}\boldsymbol{X}\boldsymbol{H}_{2,2}^H\right| \geq R_2^\star \tag{36d}$$

---
[9]Up to the inessential factor 1/2.

So far, we have proved that Problem (29) can be reformulated as in (36). This entails no loss of optimality, since no approximation has been used, yet. Nevertheless, just as (29), Problem (36) can not be solved by directly using fractional programming, because the numerator of the objective (36a) is the difference of two concave functions, and so in general is not concave. The same is true for the last constraint function, which coincides with (36a), and so the constraint set of (36) is not convex. However, (36) is in a more convenient form than (29) to exploit Proposition 4. Indeed, a concave approximation of the numerator of (36a) such that all properties (**P1**), (**P2**), and (**P3**) are fulfilled is found by considering the first-order Taylor expansion of the second summand at the numerator of (36a). To elaborate, define the concave function $g(\boldsymbol{X}) = \log_2 \left|\sigma^2 \boldsymbol{I} + \boldsymbol{H}_{2,2}\boldsymbol{X}\boldsymbol{H}_{2,2}^H\right|$. Since $g$ is concave, it can be upper-bounded by its first-order Taylor expansion at any given $\boldsymbol{X}_0 \succeq \boldsymbol{0}$ as[10]

$$g(\boldsymbol{X}) \leq \log_2\left|\sigma^2\boldsymbol{I}+\boldsymbol{H}_{2,2}\boldsymbol{X}_0\boldsymbol{H}_{2,2}^H\right|+2\Re\left\{\mathrm{tr}\left(\boldsymbol{M}_0^H(\boldsymbol{X}-\boldsymbol{X}_0)\right)\right\}$$

with $\boldsymbol{M}_0 = \boldsymbol{H}_{2,2}^H(\sigma^2\boldsymbol{I}+\boldsymbol{H}_{2,2}\boldsymbol{X}_0\boldsymbol{H}_{2,2}^H)^{-1}\boldsymbol{H}_{2,2}$ the conjugate gradient of $g$, evaluated at $\boldsymbol{X}_0$. Since the numerator of (36a) is the same as the constraint function in (36d), we can lower-bound both (36a) and (36d), thus obtaining the approximate problem

$$\max_{\boldsymbol{X}\succeq \boldsymbol{0},\boldsymbol{B}\succeq \boldsymbol{0}} \frac{\widetilde{R}_2(\boldsymbol{X},\boldsymbol{B})}{\alpha \mathrm{tr}(\boldsymbol{X}+\boldsymbol{B})+P_c} \tag{37a}$$

$$\text{s.t.} \ \frac{(c^\star+1)P_1\|\boldsymbol{H}_t\boldsymbol{h}_{1,1}\|^2\boldsymbol{h}_{2,1}^H\boldsymbol{X}\boldsymbol{h}_{2,1}}{c^\star\sigma^2\|\boldsymbol{h}_{1,1}\|^2+c^\star P_1\|\boldsymbol{H}_t\boldsymbol{h}_{1,1}\|^2} \geq \sigma^2+\boldsymbol{h}_{2,1}^H(\boldsymbol{B}+\boldsymbol{X})\boldsymbol{h}_{2,1}, \tag{37b}$$

$$\alpha \mathrm{tr}(\boldsymbol{X}+\boldsymbol{B}) \leq P_2 \ , \ \widetilde{R}_2(\boldsymbol{X},\boldsymbol{B}) \geq R_2^\star \tag{37c}$$

with

$$\widetilde{R}_2(\boldsymbol{X},\boldsymbol{B}) = \log_2\left|\sigma^2\boldsymbol{I}+\boldsymbol{H}_{2,2}(\boldsymbol{X}+\boldsymbol{B})\boldsymbol{H}_{2,2}^H\right|$$
$$- \log_2\left|\sigma^2\boldsymbol{I}+\boldsymbol{H}_{2,2}\boldsymbol{X}_0\boldsymbol{H}_{2,2}^H\right| - 2\Re\left\{\mathrm{tr}\left(\boldsymbol{M}_0^H(\boldsymbol{X}-\boldsymbol{X}_0)\right)\right\}.$$

Problem (36a) can be globally and efficiently solved by fractional programming theory, because (37a) has a concave numerator and affine denominator, and the feasible set is convex. Moreover, it is easy to verify that (37) fulfills all three properties (**P1**), (**P2**), and (**P3**) with respect to (36). Then, we can fulfill the assumptions of Proposition 4 by formulating Algorithm 2, which solves a sequence of Problems like (37), updating $\boldsymbol{X}_0$ after each iteration. Leveraging Proposition 4, Algorithm 2 monotonically increases (36a) (and hence (29)) until the objective attains a first-order optimal value.

**Remark 10.** *Algorithm 2 is implemented at the secondary transmitter, which is assumed to know the channels $\boldsymbol{H}_{2,2}$, $\boldsymbol{H}_t$, $\boldsymbol{h}_{2,1}$, and $\boldsymbol{h}_{1,1}$, and the power $P_1$. $\boldsymbol{H}_{2,2}$ and $\boldsymbol{H}_t$ can be directly estimated by the secondary system, $\boldsymbol{h}_{2,1}$ can be obtained as in the underlay case, whereas $\boldsymbol{h}_{1,1}$ can be communicated by the primary user, which is involved in the resource allocation in the overlay scenario. $P_1$ can be either communicated by the primary user, or obtained from the CQI.*

---

[10]We recall that, given a real-valued function $g$ of complex, matrix argument $\boldsymbol{X}$, the first-order Taylor expansion around a point $\boldsymbol{X}_0$ is written as $g(\boldsymbol{X}) = g(\boldsymbol{X}_0) + 2\Re\left\{\mathrm{tr}\left((\nabla_{\boldsymbol{X}^*}g)_{|\boldsymbol{X}=\boldsymbol{X}_0}^H(\boldsymbol{X}-\boldsymbol{X}_0)\right)\right\}$ [58].

**Algorithm 2** EE$_2$ maximization for overlay communications.

**Test** feasibility by (31).
**if** Feasible **then**
  $\ell = 0, \epsilon > 0$, select a feasible pair $(\boldsymbol{X}_0^{(\ell)}, \boldsymbol{B}^{(\ell)})$.
  **while** $\left|\mathrm{EE}_2\left(\boldsymbol{X}_0^{(\ell)}, \boldsymbol{B}^{(\ell)}\right) - \mathrm{EE}_2\left(\boldsymbol{X}_0^{(\ell-1)}, \boldsymbol{B}^{(\ell-1)}\right)\right| > \epsilon$ **do**
    Compute $\boldsymbol{M}_0$; Set $(\boldsymbol{X}, \boldsymbol{B})$ as the solution of Problem (37);
    $\boldsymbol{X}_0^{(\ell)} = \boldsymbol{X}; \ell = \ell + 1;$
  **end while**
**end if**

*2) Sequential fractional programming with rank-1 $\boldsymbol{A}$:* Algorithm 2 is able to operate without enforcing any restriction on $\boldsymbol{X}$ (and hence on $\boldsymbol{A}$). Instead, in order to further reduce the computational complexity, an alternative approach is to reduce the number of variables by heuristically fixing the directions of $\boldsymbol{A}$ in (29). Since it is not possible to diagonalize both the objective and the constraints, natural choices are to diagonalize either the objective (29a) or the primary user's rate $R_1$ at the left-hand side (LHS) of (29b). In the following, we opt for the latter choice, motivated by the observation that this allows one to reduce as much as possible the power devoted to relaying the primary user's message. Otherwise stated, from Proposition 3 we know that diagonalizing the LHS of (29b) maximizes the primary user's rate, and this allows us to keep $\mathrm{tr}(\boldsymbol{A})$ as low as possible while still fulfilling the rate requirement. This is anticipated to be beneficial, since the secondary user's energy efficiency is decreasing in $\mathrm{tr}(\boldsymbol{A})$.

As proved in Proposition 3, the primary user's rate is maximized when $\boldsymbol{A}$ has the structure $\boldsymbol{A} = \sqrt{a}\boldsymbol{U}_A\boldsymbol{V}_A^H$, with $\boldsymbol{U}_A = \boldsymbol{h}_{2,1}/\|\boldsymbol{h}_{2,1}\|$ and $\boldsymbol{V}_A = \boldsymbol{H}_t\boldsymbol{h}_{1,1}/\|\boldsymbol{H}_t\boldsymbol{h}_{1,1}\|$. Instead, in this case we do not fix the scalar $a$, which is left as an optimization variable. With this choice, Problem (29) can be recast as

$$\max_{a \geq 0, \boldsymbol{B} \succeq \boldsymbol{0}} \frac{R_2(\boldsymbol{A} = \sqrt{a}\boldsymbol{U}_A\boldsymbol{V}_A^H, \boldsymbol{B})}{\alpha \mathrm{tr}(\boldsymbol{B}) + \alpha a\left(P_1\frac{\|\boldsymbol{H}_t\boldsymbol{h}_{1,1}\|^2}{\|\boldsymbol{h}_{1,1}\|^2} + \sigma^2\right) + P_c} \tag{38a}$$

$$\text{s.t.} \ a \geq \frac{c^\star(\sigma^2 + \boldsymbol{h}_{2,1}^H\boldsymbol{B}\boldsymbol{h}_{2,1})\|\boldsymbol{h}_{1,1}\|^2}{\|\boldsymbol{h}_{2,1}\|^2(P_1\|\boldsymbol{H}_t\boldsymbol{h}_{1,1}\|^2 - c^\star\sigma^2\|\boldsymbol{h}_{1,1}\|^2)}, \tag{38b}$$

$$\alpha \mathrm{tr}(\boldsymbol{B}) + \alpha a \sigma^2 \psi \leq P_2 \ , \ R_2\left(\boldsymbol{A} = \sqrt{a}\boldsymbol{U}_A\boldsymbol{V}_A^H, \boldsymbol{B}\right) \geq R_2^\star \tag{38c}$$

with

$$R_2\left(\boldsymbol{A}=\sqrt{a}\boldsymbol{U}_A\boldsymbol{V}_A^H, \boldsymbol{B}\right) = -N_R\log_2(\sigma^2)$$
$$+\log_2\left|\sigma^2\boldsymbol{I}+\boldsymbol{H}_{2,2}\left[a\sigma^2\frac{\boldsymbol{h}_{2,1}\boldsymbol{h}_{2,1}^H}{\|\boldsymbol{h}_{2,1}\|^2}\psi+\boldsymbol{B}\right]\boldsymbol{H}_{2,2}^H\right|-\log_2(1+a\phi),$$

and $\psi = \left(\frac{P_1\|\boldsymbol{H}_t\boldsymbol{h}_{1,1}\|^2}{\sigma^2\|\boldsymbol{h}_{1,1}\|^2}+1\right)$, $\phi = \psi\frac{\|\boldsymbol{H}_{2,2}\boldsymbol{h}_{2,1}\|^2}{\|\boldsymbol{h}_{2,1}\|^2}$. The numerator of (38a) and the second constraint function in (38c) are again the difference of concave functions and therefore can be lower-bounded by a similar approach as in Section IV-C1, i.e. by linearizing with respect to $a$ the negative term in $R_2\left(\boldsymbol{A}=\sqrt{a}\boldsymbol{U}_A\boldsymbol{V}_A^H, \boldsymbol{B}\right)$. Then, the approximate problem

to be solved in each iteration is stated as:

$$\max_{a\geq 0, \boldsymbol{B}\succeq \boldsymbol{0}} \frac{\widehat{R_2}(a, \boldsymbol{B})}{\alpha \text{tr}(\boldsymbol{B}) + \alpha a \left(P_1 \frac{\|\boldsymbol{H}_t \boldsymbol{h}_{1,1}\|^2}{\|\boldsymbol{h}_{1,1}\|^2} + \sigma^2\right) + P_c} \quad (39a)$$

$$\text{s.t.} \quad a \geq \frac{c^\star(\sigma^2 + \boldsymbol{h}_{2,1}^H \boldsymbol{B} \boldsymbol{h}_{2,1}) \|\boldsymbol{h}_{1,1}\|^2}{\|\boldsymbol{h}_{2,1}\|^2 (P_1 \|\boldsymbol{H}_t \boldsymbol{h}_{1,1}\|^2 - c^\star \sigma^2 \|\boldsymbol{h}_{1,1}\|^2)}, \quad (39b)$$

$$\alpha \text{tr}(\boldsymbol{B}) + \alpha a \left(P_1 \|\boldsymbol{H}_t \boldsymbol{h}_{1,1}\|^2 / \|\boldsymbol{h}_{1,1}\|^2 + \sigma^2\right) \leq P_2, \quad (39c)$$

$$\widehat{R_2}(a, \boldsymbol{B}) \geq R_2^\star, \quad (39d)$$

with

$$\widehat{R_2}(a, \boldsymbol{B}) = \log_2 \left| \sigma^2 \boldsymbol{I} + \boldsymbol{H}_{2,2} \left[ a \sigma^2 \frac{\boldsymbol{h}_{2,1} \boldsymbol{h}_{2,1}^H}{\|\boldsymbol{h}_{2,1}\|^2} \psi + \boldsymbol{B} \right] \boldsymbol{H}_{2,2}^H \right|$$
$$- \log_2(1 + a_0 \phi) - \frac{\phi}{\ln(2)} \frac{a - a_0}{1 + a_0 \phi} - N_R \log_2(\sigma^2).$$

For any $a_0$, $\widehat{R_2}(a, \boldsymbol{B})$ is a concave function, and thus (39a) can be globally and efficiently solved by means of fractional programming tools. The resulting resource allocation algorithm is formulated as in Algorithm 3. With respect to Algorithm 2, Algorithm 3 is computationally simpler, since less variables are optimized, but it is more heuristic, since (38) is a suboptimal formulation of (29).

---

**Algorithm 3** Overlay. $\text{EE}_2$ maximization with rank 1 AF matrix.

**Test** feasibility by (31).
**if** Feasible **then**
  $\ell = 0$, $\epsilon > 0$, select a feasible pair $(a_0^{(\ell)}, \boldsymbol{B}^{(\ell)})$.
  **while** $\left| \text{EE}_2\left(a_0^{(\ell)}, \boldsymbol{B}^{(\ell)}\right) - \text{EE}_2\left(a_0^{(\ell-1)}, \boldsymbol{B}^{(\ell-1)}\right) \right| > \epsilon$
  **do**
    Compute $\phi$; Set $(a, \boldsymbol{B})$ as the solution of Problem (39);
    $a_0^{(\ell)} = a$; $\ell = \ell + 1$;
  **end while**
**end if**

---

## V. Numerical Simulations

Let us consider a cellular system in which a D2D communication is activated among a pair of devices which reuse the resource block dedicated to a regular cellular downlink communication. The device link is the secondary link, while the cellular link is the primary link. We set $N_{T,1} = N_{T,2} = N_R = 2$, and[11] $\alpha = 10$, $P_c = 1\,\text{W}$. Unless otherwise specified, we do not consider out-of-system interference, thus setting $\sigma^2 = FB\mathcal{N}_0$, wherein $F = 3\,\text{dB}$ is the receiver noise figure, $B = 180\,\text{kHz}$ is the communication bandwidth, and $\mathcal{N}_0 = -174\,\text{dBm/Hz}$ is the noise power spectral density. All channels are generated according to the Rayleigh fading model with path-loss model as in [61] with power decay factor of $\eta = 3.5$, and log-normal shadowing with standard deviation equal to $6\,\text{dB}$. The area to serve has a radius of $500\,\text{m}$, with the access point at the center, and the mobile terminal randomly placed within the cell, with a minimum distance from the access point of $10\,\text{m}$. The secondary terminals are also randomly placed in the cell, with a distance between them randomly generated between $10\,\text{m}$ and $100\,\text{m}$. The primary link transmits with power[12] $P_1 = -10\,\text{dBW}$, and requires a minimum rate $R_1^\star$ set as a percentage $R$ of the primary point-to-point capacity. The presented results have been obtained by averaging over 1000 independent channel realizations and terminals drops.

We begin by analyzing the performance of Algorithm 1 for the underlay scenario. Figs. 1a and 1b consider $R = 50\%, 75\%, 100\%$, and for each value of $R$ illustrate the performance of: 1) Algorithm 1 for energy efficiency optimization; 2) Algorithm 1 specialized to perform rate optimization as explained in Remark 7. For these two resource allocation schemes, Fig. 1a shows the average secondary energy efficiency versus $P_2$, whereas Fig. 1b shows the average secondary rate versus $P_2$. The results indicate that energy efficiency maximization is equivalent to rate maximization for $P_2 \leq -12\,\text{dBW}$. In this range of $P_2$, both performance functions are increasing in the transmit power. Instead, for higher $P_2$ different behaviors are observed. Increasing $P_2$ allows one to attain the peak of the energy efficiency, which then saturates as a function of $P_2$, as can be seen in Fig. 1a. Instead, the energy efficiency obtained by rate maximization decreases, because in this case the excess power is fully used to maximize the rate. This behavior is confirmed in Fig. 1b, which shows how the rate obtained by rate maximization is indeed monotonically increasing in $P_2$. Similarly, as illustrated in Fig. 2 for the case $R = 75\%$, the transmit power of the secondary system, i.e. the quantity $\text{tr}(\boldsymbol{K}_{2,1} + \boldsymbol{K}_{2,2})$, increases with $P_2$, when rate optimization is performed, whereas it eventually saturates when the energy efficiency is optimized. Moreover, we notice that the transmit power is always lower than the secondary link maximum feasible power $P_2$, due to the constraint on the primary rate requirement.

Next, we focus on the overlay scenario, recalling that this protocol is required when the primary system needs the help of the secondary transmitter to meet its own rate constraint. A typical situation where this is the case is when the primary transmitter experiences poor propagation conditions, whereas the secondary transmitter has a stronger channel to the primary receiver. Another situation where the overlay approach might prove useful is when the primary transmitter wishes to save energy by using a lower transmit power. Accordingly, for the overlay case we consider a similar scenario as in the underlay case, but we place the secondary transmitter closer to the primary receiver than the primary transmitter. In particular, users' positions are randomly generated, but with the requirement that the distance between the secondary transmitter and the primary receiver is between $10\%$ and $90\%$ of the distance between the primary transmitter and primary receiver. Similar path-loss, shadowing, and small-scale fading models are considered, setting $P_1 = -20\,\text{dBW}$, and averaging the results over 1000 independent channel and users drops

---

[11] As for $\alpha$, typical values for base stations are $3 - 4$ [59]. A higher value is considered here to account for the fact that a mobile node might be equipped with lower quality amplifiers. As for $P_c$ typical values for each transmit/receive chain are $10 - 100\,\text{dBm}$ [3], [60]. Here, the secondary system has four transmit/receive chains. Moreover a slightly higher value is considered as a conservative choice.

[12] A higher power level is normally used by base stations in present communication systems. Here, a lower power level is considered motivated by the need of reducing the energy consumptions in future networks.

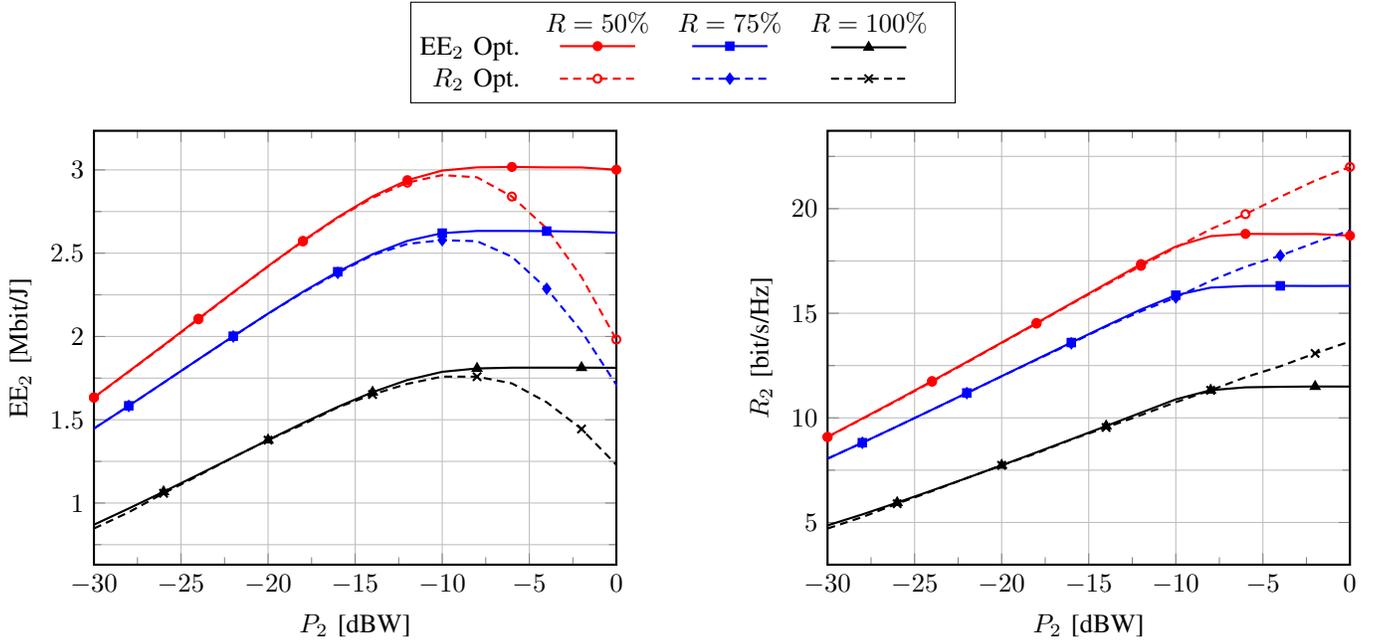

(a) Average $EE_2$ by: 1) Algorithm 1 for $EE_2$ maximization; 2) Algorithm 1 for $R_2$ maximization

(b) Average $R_2$ by: 1) Algorithm 1 for $EE_2$ maximization; 2) Algorithm 1 for $R_2$ maximization

Fig. 1. Underlay scenario: $P_1 = -10\,\text{dBW}, P_c = 1\,\text{W}, \alpha = 10, N_{T,1} = N_{T,2} = N_R = 2, B = 180\,\text{kHz}, R = 50\%, 75\%, 100\%$.

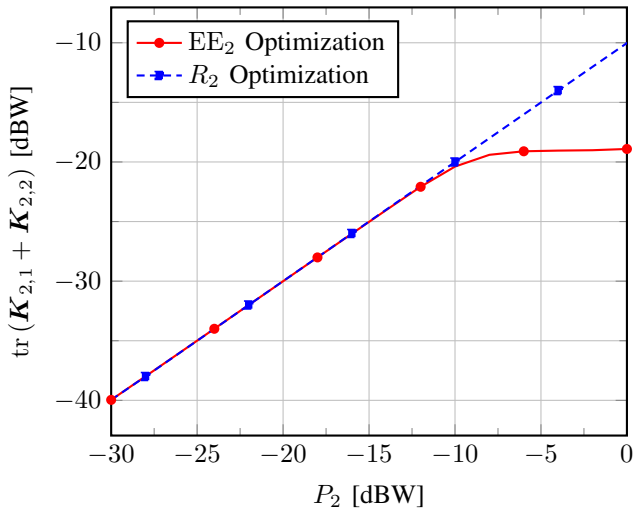

Fig. 2. Underlay scenario: $P_1 = -10\,\text{dBW}, P_c = 1\,\text{W}, \alpha = 10, N_{T,1} = N_{T,2} = N_R = 2, B = 180\,\text{kHz}, R = 75\%$. Average $\text{tr}(\boldsymbol{K}_{2,1} + \boldsymbol{K}_{2,2})$ versus $P_2$.

scenarios.

Figs. 3a and 3b are the overlay counterparts of Figs. 1a and 1b, considering $R = 125\%$, and showing the average secondary energy efficiency and rate versus $P_2$ for: 1) Algorithm 2 for energy efficiency optimization; 2) Algorithm 2 specialized to perform rate optimization; 3) Algorithm 3 for energy efficiency optimization. Similar considerations as for Figs. 1a and 1b hold. In addition, remarkably it is seen that Algorithm 3 performs similarly as Algorithm 2, despite the fact that it constrains $\boldsymbol{A}$ to have rank 1. In particular, it is seen that the gap between Algorithms 3 and 2 vanishes for higher $P_2$, thereby suggesting that for large transmit powers a rank-one solution is a suitable allocation. Thus, Algorithm 3 is a good choice in all situations in which computational complexity is a critical issue. In order to gain further insight on this point, we analyze the complexity of the proposed methods.

### A. Computational complexity

As for the underlay scenario, Algorithm 1 requires to solve at most two fractional problems, with concave numerator, affine denominator, and convex constraints. As described in Appendix A, this can be accomplished by solving a sequence of auxiliary convex problems, which converges with superlinear rate, regardless of the complexity of each auxiliary problem. Since each auxiliary problem is convex, its complexity is polynomial in the number of variables, which are at most[13] $N_{T,2}(N_{T,2} + 1)$, and of constraints, which are at most 5.

Instead, Algorithms 2 and 3, are iterative algorithms which require to solve one fractional problem in each iteration. The complexity of each fractional program can be evaluated by a similar argument as for Algorithm 1. Using Dinkelbach's algorithm, each auxiliary problem is again convex, and has $N_{T,2}(N_{T,2} + 1)$ variables[14] plus 3 constraints for Algorithm 2, while it has $N_{T,2}(N_{T,2} + 1)/2 + 1$ variables plus 3 constraints for Algorithm 3. Finally, as for the number of outer iterations required by Algorithms 2 and 3 to converge, performing a closed-form analysis appears very challenging. For this reason we resort to a numerical analysis, reporting in Tab. I the average number of iterations required for the two algorithms to reach convergence for $R = 125\%$ and $R = 200\%$. We can see that both Algorithms 2 and 3 converge in a handful of iterations. Thus, in light of the above complexity analysis, we can argue that both Algorithms 2 and 3 lend themselves to being implemented in practical networks, with Algorithm 3 enjoying an even lower complexity than Algorithm 2.

---

[13] In the underlay case we have at most two $N_{T,2} \times N_{T,2}$ Hermitian matrix variables. Therefore each matrix has $N_{T,2}(N_{T,2} + 1)/2$ free variables.

[14] Recall that although the matrix $\boldsymbol{A}$ is in general not Hermitian, we have replaced it with the Hermitian matrix $\boldsymbol{X}$.

TABLE I
AVERAGE NUMBER OF ITERATIONS FOR CONVERGENCE IN THE OVERLAY SCENARIO WITH $P_1 = -20\,\mathrm{dBW}$, $P_c = 1\,\mathrm{W}$, $\alpha = 10$, $N_{T,1} = N_{T,2} = N_R = 2$, $B = 180\,\mathrm{kHz}$, $R = 125\%, 200\%$.

| $R$ | Alg. | $P_2$ [dBW] | | | | | | | |
|---|---|---|---|---|---|---|---|---|---|
| | | -30 | -26 | -22 | -18 | -14 | -10 | -6 | -2 |
| 125% | 2 | 4.07 | 3.74 | 3.38 | 3.24 | 3.18 | 3.71 | 4.25 | 5.22 |
| | 3 | 3.39 | 2.94 | 2.69 | 2.36 | 2.32 | 2.65 | 3.45 | 4.50 |
| 200% | 2 | 4.94 | 4.19 | 4.03 | 3.56 | 3.73 | 4.27 | 4.54 | 5.64 |
| | 3 | 3.95 | 3.24 | 2.81 | 2.35 | 2.30 | 2.69 | 3.40 | 4.45 |

## VI. CONCLUSIONS

This paper has developed energy-efficient resource allocation algorithms for spectrum sharing systems following the underlay and overlay paradigms. In the underlay approach, the optimal resource allocation policy for the secondary system has been found, by means of a suitable reformulation of the original non-convex fractional problem. As for the overlay scenario, two algorithms have been proposed, which trade-off complexity with optimality claims. Numerical results indicate that the proposed schemes are effective in real-world scenarios both in terms of achieved energy efficiency and in terms of computational complexity. In particular, the two proposed algorithms for the overlay scenario perform very similarly, although the latter has weaker optimality properties.

## APPENDIX A
## FUNDAMENTALS OF FRACTIONAL PROGRAMMING

This section collects the basic results of fractional programming theory. For a more extensive overview, we refer to [3].

**Definition 1** (Fractional program). *Let $\mathcal{C} \subseteq \mathbb{R}^n$ and consider the functions $f : \mathcal{C} \to \mathbb{R}_0^+$ and $g : \mathcal{C} \to \mathbb{R}^+$. A fractional program is the optimization problem*

$$\max_{\boldsymbol{x} \in \mathcal{C}} \frac{f(\boldsymbol{x})}{g(\boldsymbol{x})} \tag{40}$$

**Proposition 6** ([62]). *An $\boldsymbol{x}^* \in \mathcal{C}$ solves (40) if and only if $\boldsymbol{x}^* = \arg\max_{\boldsymbol{x} \in \mathcal{C}} \{f(\boldsymbol{x}) - \lambda^* g(\boldsymbol{x})\}$, with $\lambda^*$ being the unique zero of $F(\lambda) = \max_{\boldsymbol{x} \in \mathcal{C}} \{f(\boldsymbol{x}) - \lambda g(\boldsymbol{x})\}$.*

This result allows us to solve (40) by finding the zero of $F(\lambda)$. To this end, the most widely used algorithm is Dinkelbach's algorithm [62], which is reported next for the reader's convenience. We can see that if $f$ and $g$ are respectively concave and convex, and if $\mathcal{C}$ is a convex set, Dinkelbach's algorithm only requires solving convex problems. Moreover, it can be shown that the update rule for $\lambda$ follows Newton's method applied to the auxiliary function $F(\lambda)$. This implies that Dinkelbach's algorithm exhibits a super-linear convergence rate. Moreover, it should be stressed that no assumption on the concavity/convexity of $f$ and $g$ has been made in Proposition 6, which then also holds if $f$ is not concave and/or $g$ is not convex and if $\mathcal{C}$ is not defined by convex constraints. In these cases, it is still possible to solve (40) using Dinkelbach's algorithm, but the auxiliary problem to be solved to compute $\boldsymbol{x}_n^*$ will be non-convex. This typically results in a too high complexity, although in some particular cases the structure of the problem allows the computation of $\boldsymbol{x}_n^*$ with affordable complexity, despite the non-convexity of the problem.

---

**Algorithm 4** Dinkelbach's algorithm
1: Initialize $\lambda_0$ with $F(\lambda_0) \geq 0$, $n = 0$;
2: **while** $F(\lambda_n) > \epsilon$ **do**
3: $\quad \boldsymbol{x}_n^* = \arg\max_{\boldsymbol{x} \in \mathcal{C}} f(\boldsymbol{x}) - \lambda_n g(\boldsymbol{x})$;
4: $\quad \lambda_{n+1} = \frac{f(\boldsymbol{x}_n^*)}{g(\boldsymbol{x}_n^*)}$; $n++$;
5: **end while**
6: Output $\boldsymbol{x}_n^*$, $\lambda_n$

---

## APPENDIX B
## PROOF OF PROPOSITION 2

Before presenting the actual proof of Proposition 2, the following lemma is shown.

**Lemma 2.** *Denote by $(\boldsymbol{K}_{2,1}^\star, \boldsymbol{K}_{2,2}^\star)$ and $(\widetilde{\boldsymbol{K}}_{2,1}, \widetilde{\boldsymbol{K}}_{2,2})$ the optimal solution to (15) and (20), respectively. If $R_{1,2}(\widetilde{\boldsymbol{K}}_{2,2}) < R_1^\star$, then $R_{1,2}(\boldsymbol{K}_{2,2}^\star) = R_1^\star$.*

In words, this lemma proves that if the solution of the relaxed problem (20) is not feasible for the original problem, then any solution of the original problem will fulfill the primary rate constraint at the secondary receiver with equality.

*Proof:* The proof follows by contradiction. Assume $R_{1,2}(\widetilde{\boldsymbol{K}}_{2,2}) < R_1^\star$ but $R_{1,2}(\boldsymbol{K}_{2,2}^\star) > R_1^\star$. Then, $(\boldsymbol{K}_{2,1}^\star, \boldsymbol{K}_{2,2}^\star)$ will also be a solution of (20). To see this, observe that if $R_{1,2}(\boldsymbol{K}_{2,2}^\star) > R_1^\star$ then the Lagrange multiplier associated with this constraint will be zero by virtue of the complementary slackness condition. This implies that the KKT conditions satisfied by $(\boldsymbol{K}_{2,1}^\star, \boldsymbol{K}_{2,2}^\star)$ are formally equal to the KKT conditions of (20), which are necessary and sufficient for optimality since the problem is a concave fractional problem. Moreover, by virtue of Proposition 1 and Lemma 1, (20) is equivalent to (19) and has the unique solution $(\widetilde{\boldsymbol{K}}_{2,1}, \widetilde{\boldsymbol{K}}_{2,2}) = (\boldsymbol{0}, \boldsymbol{\Sigma}^\star)$. As a consequence, $\boldsymbol{K}_{2,2}^\star = \widetilde{\boldsymbol{K}}_{2,2} = \boldsymbol{\Sigma}^\star$ which implies $R_{1,2}(\widetilde{\boldsymbol{K}}_{2,2}) = R_{1,2}(\boldsymbol{K}_{2,2}^\star) < R_1^\star$. This is a contradiction since we have assumed $R_{1,2}(\boldsymbol{K}_{2,2}^\star) > R_1^\star$. ■

Exploiting Lemma 2, Proposition 2 can be proved as follows.

*Proof:* Given (21), the solution of the relaxed problem (20) will not be feasible for (15). Indeed, putting together the results of Proposition 1 and Lemma 1, it follows that Problem (20) has the unique solution $(\boldsymbol{K}_{2,1}^\star = \boldsymbol{0}, \boldsymbol{K}_{2,2}^\star = \boldsymbol{\Sigma}^\star)$, which can not be feasible for (15) due to (21). Then, by virtue of

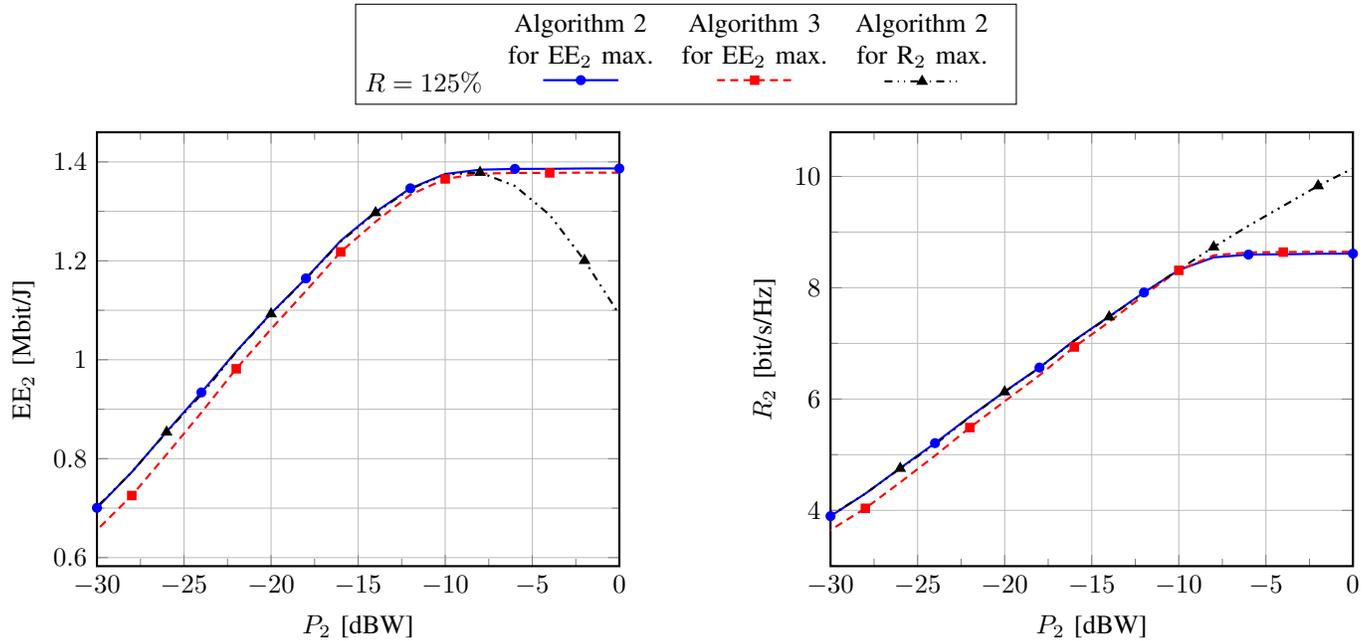

(a) Average $EE_2$ by: 1) Algorithm 2 for $EE_2$ maximization; 2) Algorithm 3 for $EE_2$ maximization; 3) Algorithm 2 for $R_2$ maximization.

(b) Average $R_2$ by: 1) Algorithm 2 for $EE_2$ maximization; 2) Algorithm 3 for $EE_2$ maximization; 3) Algorithm 2 for $R_2$ maximization.

Fig. 3. Overlay scenario: $P_1 = -20\,\text{dBW}, P_c = 1\,\text{W}, \alpha = 10, N_{T,1} = N_{T,2} = N_R = 2, B = 180\,\text{kHz}, R = 200\%$.

Lemma 2, Problem (15) can be equivalently written as

$$\max_{\substack{\boldsymbol{K}_{2,1} \succeq \boldsymbol{0} \\ \boldsymbol{K}_{2,2} \succeq \boldsymbol{0}}} \frac{\log_2\left|\boldsymbol{I}+\frac{1}{\sigma^2}\boldsymbol{H}_{2,2}(\boldsymbol{K}_{2,1}+\boldsymbol{K}_{2,2})\boldsymbol{H}_{2,2}^H+\frac{1}{\sigma^2}\boldsymbol{Q}_1\right|-R_1^\star}{\alpha\text{tr}(\boldsymbol{K}_{2,1}+\boldsymbol{K}_{2,2})+P_c} \quad (41\text{a})$$

s.t. $\boldsymbol{h}_{2,1}^H(\boldsymbol{K}_{2,1}+\boldsymbol{K}_{2,2})\boldsymbol{h}_{2,1} \leq P_{\text{int}}$, $\alpha\text{tr}(\boldsymbol{K}_{2,1}+\boldsymbol{K}_{2,2}) \leq P_2$ (41b)

$\log_2\left|\boldsymbol{I}+\boldsymbol{Q}_1(\sigma^2\boldsymbol{I}+\boldsymbol{H}_{22}\boldsymbol{K}_{2,2}\boldsymbol{H}_{22}^H)^{-1}\right| = R_1^\star$ (41c)

$\log_2\left|\boldsymbol{I}+\frac{1}{\sigma^2}\boldsymbol{H}_{2,2}(\boldsymbol{K}_{2,1}+\boldsymbol{K}_{2,2})\boldsymbol{H}_{2,2}^H+\frac{1}{\sigma^2}\boldsymbol{Q}_1\right|-R_1^\star \geq R_2^\star$, (41d)

with $P_{\text{int}}$ as defined in (16). The next step is to observe that Problem (24) is a relaxed version of (41) obtained by transforming the equality constraint (41c) into an inequality constraint. At this point, let us denote by $(\widehat{\boldsymbol{K}}_{2,1}, \widehat{\boldsymbol{K}}_{2,2})$ a solution of (24). Then it is always possible to construct the matrices $\boldsymbol{K}_{2,1} = \widehat{\boldsymbol{K}}_{2,1} + (1-\gamma)\widehat{\boldsymbol{K}}_{2,2}$ and $\boldsymbol{K}_{2,2} = \gamma\widehat{\boldsymbol{K}}_{2,2}$, where $\gamma \in [0,1]$. Thus, for any $\gamma \in [0,1]$, $\boldsymbol{K}_{2,1} + \boldsymbol{K}_{2,2} = \widehat{\boldsymbol{K}}_{2,1} + \widehat{\boldsymbol{K}}_{2,2}$, which implies that the pair $(\boldsymbol{K}_{2,1}, \boldsymbol{K}_{2,2})$ yields the same value of the objective as $(\widehat{\boldsymbol{K}}_{2,1}, \widehat{\boldsymbol{K}}_{2,2})$ and fulfills Constraints (41b) and (41d). Moreover, when $\gamma = 0$, it holds $R_{1,2} = \log_2\left(1+\frac{P_1\|\boldsymbol{H}_{1,2}\boldsymbol{h}_{1,1}\|^2}{\sigma^2\|\boldsymbol{h}_{1,1}\|^2}\right) > R_1^\star$ due to (21); when $\gamma = 1$, it holds $R_{1,2} = \log_2\left|\boldsymbol{I}+\boldsymbol{Q}_1(\sigma^2\boldsymbol{I}+\boldsymbol{H}_{22}\boldsymbol{K}_{2,2}\boldsymbol{H}_{22}^H)^{-1}\right| \leq R_1^\star$, due to (24c). Then, it is always possible to find $\gamma \in [0;1]$ such that (41c) holds, i.e. $R_{1,2} = R_1^\star$. Thus, for this choice of $\gamma$, the pair $(\boldsymbol{K}_{2,1}, \boldsymbol{K}_{2,2})$ that we have constructed is not only a solution of the relaxed problem (24), but is also feasible for Problem (41). This implies that $(\boldsymbol{K}_{2,1}, \boldsymbol{K}_{2,2})$ is also a solution of (41), and hence of (15) by virtue of Lemma 2. ∎

## APPENDIX C
### PROOF OF PROPOSITION 3

*Proof:* Let $\boldsymbol{A} = \boldsymbol{U}_A \boldsymbol{\Lambda}_A^{1/2} \boldsymbol{V}_A^H$ be the singular value decomposition (SVD) of $\boldsymbol{A}$. From (28) we can see that for any $\boldsymbol{A}$, $R_1$ is maximized with respect to $\boldsymbol{B}$ when[15] $\boldsymbol{B} = \boldsymbol{0}$. Next, the rate-maximizing $\boldsymbol{A}$ is found as the solution to the following optimization problem

$$\max_{\boldsymbol{A}} \frac{P_1}{\|\boldsymbol{h}_{1,1}\|^2} \frac{|\boldsymbol{h}_{2,1}^H \boldsymbol{A} \boldsymbol{H}_t \boldsymbol{h}_{1,1}|^2}{\sigma^2 + \sigma^2 \boldsymbol{h}_{2,1}^H \boldsymbol{A} \boldsymbol{A}^H \boldsymbol{h}_{2,1}}, \text{ s.t. } \alpha\text{tr}(\boldsymbol{A}\boldsymbol{M}\boldsymbol{A}^H) \leq P_2, \quad (42)$$

which, leveraging the results in Appendix A, is equivalent to:

$$\max_{\boldsymbol{A}} \frac{P_1}{\|\boldsymbol{h}_{1,1}\|^2} |\boldsymbol{h}_{2,1}^H \boldsymbol{A} \boldsymbol{H}_t \boldsymbol{h}_{1,1}|^2 - \mu\sigma^2\left(1+\boldsymbol{h}_{2,1}^H \boldsymbol{A}\boldsymbol{A}^H \boldsymbol{h}_{2,1}\right) \quad (43\text{a})$$

s.t $\alpha\text{tr}(\boldsymbol{A}\boldsymbol{M}\boldsymbol{A}^H) \leq P_2$, (43b)

for the specific choice of $\mu \geq 0$ such that the maximum of (43a) is equal to zero. Elaborating, for any $\mu \geq 0$, (43a) can be rewritten as $\boldsymbol{h}_{2,1}^H \boldsymbol{A}(\boldsymbol{M} - \sigma^2(\mu+1)\boldsymbol{I}_{N_{T,2}})\boldsymbol{A}^H \boldsymbol{h}_{2,1} - \mu\sigma^2$. Thus, the optimal $\boldsymbol{A}$ is such that $\boldsymbol{U}_A = \boldsymbol{h}_{2,1}$ and $\boldsymbol{V}_A = \boldsymbol{U}_M$, with $\boldsymbol{U}_M$ the eigenvector matrix of $\boldsymbol{M}$, so as to diagonalize the channel $\boldsymbol{h}_{2,1}$ and the matrix $\boldsymbol{M}$. Next, recalling that $\boldsymbol{M} = \frac{P_1}{\|\boldsymbol{h}_{1,1}\|^2} \boldsymbol{H}_t \boldsymbol{h}_{1,1} \boldsymbol{h}_{1,1}^H \boldsymbol{H}_t^H + \sigma^2 \boldsymbol{I}_{N_{T,2}}$, and observing that $\frac{P_1}{\|\boldsymbol{h}_{1,1}\|^2} \boldsymbol{H}_t \boldsymbol{h}_{1,1} \boldsymbol{h}_{1,1}^H \boldsymbol{H}_t^H$ is a rank-one matrix with non-zero eigenvalue equal to $P_1\|\boldsymbol{H}_t\boldsymbol{h}_{1,1}\|^2/\|\boldsymbol{h}_{1,1}\|^2$, we see that the largest eigenvalue of the matrix $\boldsymbol{M} - \sigma^2(\mu+1)\boldsymbol{I}_{N_{T,2}}$ is $P_1\|\boldsymbol{H}_t\boldsymbol{h}_{1,1}\|^2/\|\boldsymbol{h}_{1,1}\|^2 - \sigma^2\mu$, which corresponds to the eigenvector $\boldsymbol{H}_t\boldsymbol{h}_{1,1}/\|\boldsymbol{H}_t\boldsymbol{h}_{1,1}\|$. Instead, all other eigenvalues are negative and equal to $-\sigma^2\mu$. Since it is suboptimal to al-

---

[15]$\boldsymbol{h}_{2,1}^H \boldsymbol{B} \boldsymbol{h}_{2,1} = 0$ also if $\boldsymbol{B} \succ \boldsymbol{0}$ and $\boldsymbol{h}_{2,1}$ is in the null space of $\boldsymbol{B}$. However, this would waste power for $\boldsymbol{B}$, restricting the feasible set for $\boldsymbol{A}$.

locate power along the eigenvectors of $M$ corresponding to negative eigenvalues[16], the optimal $\mathbf{\Lambda}_A$ must be such that $\mathbf{\Lambda}_A = [a, 0, \ldots, 0]$, with $a$ a positive scalar. Plugging the optimal $\mathbf{U}_A$, $\mathbf{V}_A$, and $\mathbf{\Lambda}_A$ into (42), and since the objective is increasing in $a$, the optimal $a$ fulfills the power constraint in (42) with equality. Hence the result. ∎

---

[16]Otherwise we could increase (43a) without violating (43b), setting to zero the singular values of $A$ corresponding to the negative eigenvalues of $M$.

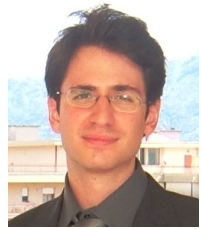

**Alessio Zappone** (S'08 – M'11 – SM'16) received his Ph.D. from the University of Cassino and Southern Lazio, Cassino, Italy. Afterwards, he worked with Consorzio Nazionale Interuniversitario per le Telecomunicazioni (CNIT) in the framework of the FP7 EU-funded project TREND. From 2012 to 2016, Alessio has been with the Department of Communication Theory of the Technische Universität Dresden, Dresden, Germany, serving as the principal investigator of the project CEMRIN, funded by the German research foundation (DFG) and carried out at the Department of Communication Theory of the Technische Universität Dresden, Dresden, Germany. Since 2016 he is adjunct professor with the University of Cassino and Southern Lazio.

His research interests lie in the area of communication theory and signal processing, with main focus on optimization techniques for resource allocation and energy efficiency. He held several research appointments at TU Dresden, Politecnico di Torino, Suplec - Alcatel-Lucent Chair on Flexible Radio, and University of Naples Federico II. He was the recipient of a Newcom# mobility grant in 2014. Alessio serves as associate editor for the IEEE SIGNAL PROCESSING LETTERS and has served as associate editor for the IEEE JOURNAL ON SELECTED AREAS ON COMMUNICATIONS (Special Issue on Energy-Efficient Techniques for 5G Wireless Communication Systems).

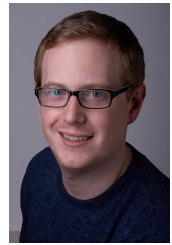

**Bho Matthiesen** (S'12) received the Diplom-Ingenieur (M.S.) degree in electrical engineering from the Technische Universität Dresden, Germany, in 2006. Since May 2012, he has been a Research Assistant with the Chair of Communications Theory at the Technische Universität Dresden, Germany, where he is currently pursuing his Ph.D. degree. His research interests are in the area of information theory, resource allocation, and communications theory.

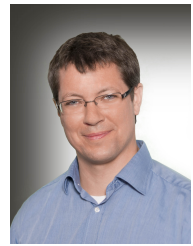

**Eduard Jorswieck** (S'01 – M'03 – SM'08) received the Diplom-Ingenieur (M.S.) degree and Doktor-Ingenieur (Ph.D.) degree, both in electrical engineering and computer science, from the Technische Universität Berlin, Germany, in 2000 and 2004, respectively. He was with the Broadband Mobile Communication Networks Department, Fraunhofer Institute for Telecommunications, Heinrich-Hertz-Institut, Berlin, from 2000 to 2008. From 2005 to 2008, he was a Lecturer with the Technische Universität Berlin. From 2006 to 2008, he was with the Department of Signals, Sensors and Systems, Royal Institute of Technology, as a Post-Doctoral Researcher and an Assistant Professor. Since 2008, he has been the Head of the Chair of Communications Theory and a Full Professor with the Technische Universität Dresden, Germany. He is principal investigator in the excellence cluster center for Advancing Electronics Dresden (cfAED) and founding member of the 5G lab Germany (5Glab.de).

His main research interests are in the area of signal processing for communications and networks, applied information theory, and communications theory. He has authored over 80 journal papers, 8 book chapters, some 225 conference papers and 3 monographs on these research topics. Eduard was a co-recipient of the IEEE Signal Processing Society Best Paper Award in 2006 and co-authored papers that won the Best Paper or Best Student Paper Awards at IEEE WPMC 2002, Chinacom 2010, IEEE CAMSAP 2011, IEEE SPAWC 2012, and IEEE WCSP 2012.

Dr. Jorswieck was a member of the IEEE SPCOM Technical Committee (2008 - 2013), and has been a member of the IEEE SAM Technical Committee since 2015. Since 2011, he has been an Associate Editor of the IEEE TRANSACTIONS ON SIGNAL PROCESSING. Since 2008, continuing until 2011, he has served as an Associate Editor of the IEEE SIGNAL PROCESSING LETTERS, and until 2013, as a Senior Associate Editor. Since 2013, he has served as an Editor of the IEEE TRANSACTIONS ON WIRELESS COMMUNICATIONS.